\documentclass[12pt,titlepage,letterpaper]{utarticle}

\usepackage{amsmath}
\usepackage{mathrsfs}
\usepackage{amsfonts}
\usepackage{amssymb}
\usepackage{amsthm}
\usepackage{mathtools}
\usepackage{mathbbol}
\usepackage{graphicx}
\usepackage{color}
\usepackage{ucs}
\usepackage[utf8x]{inputenc}
\usepackage{xparse}
\usepackage{tocloft}
\usepackage[titletoc,title]{appendix}
\usepackage{cite}
\usepackage{hyperref}
\usepackage{longtable}

\definecolor{aqua}{rgb}{0, 1.0, 1.0}
\definecolor{fuschia}{rgb}{1.0, 0, 1.0}
\definecolor{gray}{rgb}{0.502, 0.502, 0.502}
\definecolor{lime}{rgb}{0, 1.0, 0}
\definecolor{maroon}{rgb}{0.502, 0, 0}
\definecolor{navy}{rgb}{0, 0, 0.502}
\definecolor{olive}{rgb}{0.502, 0.502, 0}
\definecolor{purple}{rgb}{0.502, 0, 0.502}
\definecolor{silver}{rgb}{0.753, 0.753, 0.753}
\definecolor{teal}{rgb}{0, 0.502, 0.502}

\makeatletter
\newdimen\itex@wd%
\newdimen\itex@dp%
\newdimen\itex@thd%
\def\itexspace#1#2#3{\itex@wd=#3em%
\itex@wd=0.1\itex@wd%
\itex@dp=#2ex%
\itex@dp=0.1\itex@dp%
\itex@thd=#1ex%
\itex@thd=0.1\itex@thd%
\advance\itex@thd\the\itex@dp%
\makebox[\the\itex@wd]{\rule[-\the\itex@dp]{0cm}{\the\itex@thd}}}
\makeatother

\makeatletter
\newif\if@sup
\newtoks\@sups
\def\append@sup#1{\edef\act{\noexpand\@sups={\the\@sups #1}}\act}%
\def\reset@sup{\@supfalse\@sups={}}%
\def\mk@scripts#1#2{\if #2/ \if@sup ^{\the\@sups}\fi \else%
  \ifx #1_ \if@sup ^{\the\@sups}\reset@sup \fi {}_{#2}%
  \else \append@sup#2 \@suptrue \fi%
  \expandafter\mk@scripts\fi}
\def\tensor#1#2{\reset@sup#1\mk@scripts#2_/}
\def\multiscripts#1#2#3{\reset@sup{}\mk@scripts#1_/#2%
  \reset@sup\mk@scripts#3_/}
\makeatother

\makeatletter
\newbox\slashbox \setbox\slashbox=\hbox{$/$}
\def\itex@pslash#1{\setbox\@tempboxa=\hbox{$#1$}
  \@tempdima=0.5\wd\slashbox \advance\@tempdima 0.5\wd\@tempboxa
  \copy\slashbox \kern-\@tempdima \box\@tempboxa}
\def\slash{\protect\itex@pslash}
\makeatother

\def\clap#1{\hbox to 0pt{\hss#1\hss}}

\let\oldroot\root
\def\root#1#2{\oldroot #1 \of{#2}}
\renewcommand{\sqrt}[2][]{\oldroot #1 \of{#2}}

\DeclareSymbolFont{symbolsC}{U}{txsyc}{m}{n}
\SetSymbolFont{symbolsC}{bold}{U}{txsyc}{bx}{n}
\DeclareFontSubstitution{U}{txsyc}{m}{n}

\DeclareSymbolFont{stmry}{U}{stmry}{m}{n}
\SetSymbolFont{stmry}{bold}{U}{stmry}{b}{n}

\DeclareFontFamily{OMX}{MnSymbolE}{}
\DeclareSymbolFont{mnomx}{OMX}{MnSymbolE}{m}{n}
\SetSymbolFont{mnomx}{bold}{OMX}{MnSymbolE}{b}{n}
\DeclareFontShape{OMX}{MnSymbolE}{m}{n}{
    <-6>  MnSymbolE5
   <6-7>  MnSymbolE6
   <7-8>  MnSymbolE7
   <8-9>  MnSymbolE8
   <9-10> MnSymbolE9
  <10-12> MnSymbolE10
  <12->   MnSymbolE12}{}

\makeatletter
\def\re@DeclareMathSymbol#1#2#3#4{%
    \let#1=\undefined
    \DeclareMathSymbol{#1}{#2}{#3}{#4}}
\re@DeclareMathSymbol{\neArrow}{\mathrel}{symbolsC}{116}
\re@DeclareMathSymbol{\neArr}{\mathrel}{symbolsC}{116}
\re@DeclareMathSymbol{\seArrow}{\mathrel}{symbolsC}{117}
\re@DeclareMathSymbol{\seArr}{\mathrel}{symbolsC}{117}
\re@DeclareMathSymbol{\nwArrow}{\mathrel}{symbolsC}{118}
\re@DeclareMathSymbol{\nwArr}{\mathrel}{symbolsC}{118}
\re@DeclareMathSymbol{\swArrow}{\mathrel}{symbolsC}{119}
\re@DeclareMathSymbol{\swArr}{\mathrel}{symbolsC}{119}
\re@DeclareMathSymbol{\nequiv}{\mathrel}{symbolsC}{46}
\re@DeclareMathSymbol{\Perp}{\mathrel}{symbolsC}{121}
\re@DeclareMathSymbol{\Vbar}{\mathrel}{symbolsC}{121}
\re@DeclareMathSymbol{\sslash}{\mathrel}{stmry}{12}
\re@DeclareMathSymbol{\bigsqcap}{\mathop}{stmry}{"64}
\re@DeclareMathSymbol{\biginterleave}{\mathop}{stmry}{"6}
\re@DeclareMathSymbol{\invamp}{\mathrel}{symbolsC}{77}
\re@DeclareMathSymbol{\parr}{\mathrel}{symbolsC}{77}
\makeatother

\makeatletter
\def\Decl@Mn@Delim#1#2#3#4{%
  \if\relax\noexpand#1%
    \let#1\undefined
  \fi
  \DeclareMathDelimiter{#1}{#2}{#3}{#4}{#3}{#4}}
\def\Decl@Mn@Open#1#2#3{\Decl@Mn@Delim{#1}{\mathopen}{#2}{#3}}
\def\Decl@Mn@Close#1#2#3{\Decl@Mn@Delim{#1}{\mathclose}{#2}{#3}}
\Decl@Mn@Open{\llangle}{mnomx}{'164}
\Decl@Mn@Close{\rrangle}{mnomx}{'171}
\Decl@Mn@Open{\lmoustache}{mnomx}{'245}
\Decl@Mn@Close{\rmoustache}{mnomx}{'244}
\makeatother

\makeatletter
\DeclareRobustCommand\widecheck[1]{{\mathpalette\@widecheck{#1}}}
\def\@widecheck#1#2{%
    \setbox\z@\hbox{\m@th$#1#2$}%
    \setbox\tw@\hbox{\m@th$#1%
       \widehat{%
          \vrule\@width\z@\@height\ht\z@
          \vrule\@height\z@\@width\wd\z@}$}%
    \dp\tw@-\ht\z@
    \@tempdima\ht\z@ \advance\@tempdima2\ht\tw@ \divide\@tempdima\thr@@
    \setbox\tw@\hbox{%
       \raise\@tempdima\hbox{\scalebox{1}[-1]{\lower\@tempdima\box
\tw@}}}%
    {\ooalign{\box\tw@ \cr \box\z@}}}
\makeatother

\makeatletter
\NewDocumentCommand\mathraisebox{moom}{%
\IfNoValueTF{#2}{\def\@temp##1##2{\raisebox{#1}{$\m@th##1##2$}}}{%
\IfNoValueTF{#3}{\def\@temp##1##2{\raisebox{#1}[#2]{$\m@th##1##2$}}%
}{\def\@temp##1##2{\raisebox{#1}[#2][#3]{$\m@th##1##2$}}}}%
\mathpalette\@temp{#4}}
\makeatletter

\makeatletter
\def\udots{\mathinner{\mkern2mu\raise\p@\hbox{.}
\mkern2mu\raise4\p@\hbox{.}\mkern1mu
\raise7\p@\vbox{\kern7\p@\hbox{.}}\mkern1mu}}
\makeatother

\theoremstyle{plain}

\theoremstyle{definition}

\theoremstyle{remark}

\allowdisplaybreaks

\begin{document}

%-------------------------------------------------------------------
\preprint{
UTTG--05--18\\
%ICTP--SAIFR/2017--XX\\
}

\title{Tinkertoys for the $E_8$ Theory}

\author{Oscar Chacaltana
    \address{
    Howard Community College\\
    10901 Little Patuxent Pkwy\\
    Columbia, MD 21044, USA\\
    {~}\\
    \email{ochacaltanaalarcon@howardcc.edu}\\
    },
    Jacques Distler
     \address{
     Theory Group\\
     Department of Physics,\\
     University of Texas at Austin,\\
     Austin, TX 78712, USA \\
     {~}\\
      \email{distler@golem.ph.utexas.edu}\\
      \email{zhuyinan@physics.utexas.edu}
	},
	Anderson Trimm
	\address{
	Department of Mathematics, \\
     Gwinnett School of Mathematics,\\
     Science and Technology,\\
     970 McElvaney Lane,\\
     Lawrenceville, GA 30044, USA\\
     {~}\\
     \email{anderson.trimm@gsmst.org}
     }
     and Yinan Zhu ${}^\mathrm{b}$
}

\date{\today}
%\date{January 1, 2015}

\Abstract{
We construct the 4D $\mathcal{N}=2$ SCFTs of class-S, which stem from the $E_8$ (2,0) theory. There are 49,836 isolated SCFTs which arise as 3-punctured spheres. Of these, 149 are ``mixed" (contain free hypermultiplets accompanying the interacting SCFT) and 775 have enhanced global symmetries (beyond the manifest global symmetry associated to the punctures). Among the 49,836 3-punctured spheres we find (after removing any free hypermultiplets which may be present) 29 that are product SCFTs.  Turning to 4-punctured spheres, we find 1,025,438 4D SCFTs arising as a gauging (with a simple gauge group) of a pair of 3-punctured spheres. We discuss a number of applications, including recovering several known 4D SCFTs. Our full set of results can be accessed on the Web at  \href{https://golem.ph.utexas.edu/class-S/E8/}{https://golem.ph.utexas.edu/class-S/E8/}.
}

\maketitle

\tocloftpagestyle{empty}
\tableofcontents
\vfill
\newpage
\setcounter{page}{1}

\section{Introduction}\label{introduction}

In the Wilsonian approach, a continuum quantum field theory is a relevant perturbation of a conformal theory. A first step, then, in understanding all quantum field theories, is to understand the conformal theories. With extended supersymmetry, this seems like an achievable goal and tremendous progress has been made in recent years.

With $\mathcal{N}=2$ supersymmetry, the class-S construction \cite{Gaiotto:2009hg,Gaiotto:2009we} has proven a very fruitful tool. These theories arise as compactifications of the 6D (2,0) theory on a punctured Riemann surface, with a partial topological twist along the fibers. The punctures are the locations of (4D spacetime-filling) codimension-2 defect operators which are labeled by nilpotent orbits in the ADE Lie algebra of the (2,0) theory. The 4D SCFT, thus obtained, has a space of exactly marginal deformations which is (roughly) $\overline{\mathcal{M}}_{g,n}$, the moduli space of complex structures for the punctured Riemann surface, $C$. At the boundaries of $\mathcal{M}_{g,n}$, where $C$ develops a node, the SCFT decomposes into a product of free vector multiplets and the SCFTs corresponding to each of the components of the normalization of $C$. Moving away from the boundary corresponds to turning on a marginal (complex) gauge coupling for a weakly-gauged subgroup of the flavour symmetry of the theory. Each pants-decomposition of $C$ gives rise to a different (``S-dual'') presentation of this family of theories as a gauging of a product of SCFTs corresponding to 3-punctured spheres. To understand the theories that can arise in this way, we need to classify the SCFTs corresponding to 3-punctured spheres and their allowed gaugings. Arbitrarily complicated 4D SCFTs can then be assembled, in tinkertoy fashion.

In previous papers, we have carried out this program for the (2,0) theories of A \cite{Chacaltana:2010ks} and D \cite{Chacaltana:2011ze} types, and for $E_6$ \cite{Chacaltana:2014jba} and $E_7$ \cite{Chacaltana:2017boe} (as well as for the twisted $A_{2N-1}$ \cite{Chacaltana:2012ch}, twisted $D_N$ \cite{Chacaltana:2013oka,Chacaltana:2016shw} and twisted $E_6$ \cite{Chacaltana:2015bna} theories). Here, we complete the classification by presenting the tinkertoy catalogue for $E_8$.

There are 70 nilpotent orbits in $\mathfrak{e}_8$. Excluding the regular orbit (which corresponds to the trivial defect), that gives 69 codimension-2 defect operators (``regular punctures'') in the $E_8$ (2,0) theory. Given a triple of defects, we can compactify the (2,0) theory on a 3-punctured sphere (fixture). Naively, that would give rise to 57,155 different fixtures. Of these, 7,319 are ``bad'' (do not lead to well-defined 4D theories)\footnote{A simple diagnostic for when a fixture is bad will be given in \S\ref{computing_the_halllittlewood_superconformal_index}}. But that still leaves 49,836 new 4D SCFTs, arising as spheres with 3 regular punctures.

To these, we must supplement a set of ``irregular fixtures'' which result from the collision of a pair of punctures (conformally-equivalent to bubbling off a 3-punctured sphere) where the 3-punctured sphere that would have resulted from this degeneration of $C$ (with the third puncture being the full puncture) would have been bad. There are 50 irregular fixtures in the $E_8$ theory. These never occur in isolation; they only appear when $C$ degenerates in such a way that one of the components of its normalization has genus-0 and 6 or fewer punctures.

Unlike all of the previous cases, where there was at least one regular fixture consisting of free hypermultiplets, all of the regular fixtures in the $E_8$ theory are interacting SCFTs (or ``mixed'' fixtures, consisting of an interacting SCFT plus some number of free hypermultiplets). As a consequence, \emph{none} of the 4D SCFTs obtained via the class-S construction for $E_8$ has an interpretation as a Lagrangian field theory in \emph{any} S-duality frame.

Still, there are a vast number of them. If we consider the next-simplest case, a 4-punctured sphere, we have a gauge theory with a simple gauge group, coupled to some appropriate matter content (consisting of interacting SCFTs and perhaps some additional hypermultiplets) which ensures the vanishing of the $\beta$-function. There are 1,025,438 distinct possibilities (after eliminating the bad ones).

Since these numbers are vastly too large to list in a paper, we have created an interactive web application \href{[https://golem.ph.utexas.edu/class-S/E8/}{https://golem.ph.utexas.edu/class-S/E8/} where the interested reader can explore them.

Most of these results are a straightforward application of the ideas of \cite{Chacaltana:2012zy}, in that much of the physics is captured by the local properties of the punctures.

Three parts of the calculation are particularly arduous, though.

\begin{itemize}%
\item Determining which fixtures have enhanced global symmetries (enhanced over the manifest global symmetry group, which is the product of the global symmetries associated to each of the punctures) and which fixtures contain free hypermultiplets requires a computations of the Hall-Littlewood index \cite{Gaiotto:2012uq} to order $\tau^2$. This is tedious because the index receives contributions from $E_8$ representations of very high dimensions.
\item Determining which fixtures are product SCFTs, using the methods of \cite{Distler:2017xba} is even more arduous, as it requires computing the Hall-Littlewood index to order $\tau^4$ and the Schur index to the same order. Fortunately, we have found a simpler diagnostic which gives a sufficient (it is unknown whether it is necessary) condition for a fixture to be a product SCFT. \cite{Beem:2013sza} proved a lower bound on the level of the current algebra for any simple factor in the global symmetry group. When this bound is violated, an assumption that went into the proof --- namely that the SCFT has only a single stress-tensor multiplet --- is violated, and the theory must be a product SCFT. We found 22 fixtures which are product SCFTs and where we were able to identify the factors in the product. As discussed in \S\ref{theories_with__global_symmetry},\ref{more_product_theories_from_enhanced_}, there are 7 more fixtures that we know to be product SCFTs, but were not able to definitively identify the factors in the product. 
\item Finally, to determine the Coulomb branch geometry of any of these theories, we need the detailed form of the contraints on the pole coefficients of the meromorphic $k$-differentials which have poles at the locations of the punctures. This is, by far, the most tedious part of the calculation, and one that we have not yet completed. So far, we have only determined the constraints associated to half of the punctures (and have partial results for the rest). Nevertheless, this was enough to bootstrap our knowledge of each puncture's contribution to the graded Coulomb branch dimensions. The full set of constraints will be presented elsewhere \cite{Chacaltana:2018xxx}.
\end{itemize}

\section{The $E_8$ Theory}\label{the__theory}

\subsection{Puncture Properties}\label{puncture_properties}

As usual, we use Bala-Carter notation \cite{BalaCarter1,BalaCarter2} to label the nilpotent orbits, where $O_N=0$ is the full puncture and $O_N=E_8(a_1)$ is the simple puncture.

The flavour symmetry algebra, $\mathfrak{f}$, associated to a puncture is the centralizer of $\rho_N:\mathfrak{su}(2)\to \mathfrak{e}_8$. For the distinguished orbits, this is trivial, whereas for the $0$ orbit, it is all of $\mathfrak{e}_8$. The level of each simple factor, $\mathfrak{f}_i\subset\mathfrak{f}$ is determined from the decomposition of the adjoint representation under the embedding $\mathfrak{e}_8\supset \mathfrak{su}(2)\times \mathfrak{f}_i$. Writing this decomposition as

\begin{equation}
\mathfrak{e}_8 = \bigoplus_n V_n\otimes R_{n,i}
\label{decomp}\end{equation}
where $V_n$ is the $n$-dimensional irreducible representation of $\mathfrak{su}(2)$ and $R_{n,i}$ is the corresponding (in general, reducible) representation of $\mathfrak{f}_i$, the level of $\mathfrak{f}_i$ is given by

\begin{displaymath}
k_i = \sum_n l_{n,i}
\end{displaymath}
 where $l_{n,i}$ is the Dynkin index of the representation $R_{n,i}$. The normalization of the levels is such that vanishing $\beta$-function for a gauging of $F_i$ is achieved when the total level of all of the matter charged under $F_i$ is equal to $4\kappa_{\mathfrak{f}_i}$, where $\kappa$ is the dual Coxeter number. There are 10 distinguished punctures, whose flavour symmetry algebra is trivial.

For $\mathfrak{f}_i = \mathfrak{u}(1)$, the level is computed as above, where $l_{n,i}$ is the $\mathfrak{u}(1)$ charge squared. We normalize the $\mathfrak{u}(1)$ generators so that the free hypermultiplets in the mixed fixtures have charge 1. However, the only puncture with a $\mathfrak{u}(1)$ factor in its flavor symmetry algebra with free hypers charged under that $\mathfrak{u}(1)$ is $A_3+A_2$. Since we don't know how to normalize the other $\mathfrak{u}(1)$ generators, we don't compute their levels.

At a puncture $\Phi(z)$ has a simple pole with nilpotent residue,

\begin{displaymath}
\Phi(z) = \frac{E}{z} + \text{regular}
\end{displaymath}
where $E$ is a representative of the ``Hitchin'' Nilpotent orbit which is Spaltenstein-dual to the Nahm orbit (which we use to label our punctures)

\begin{displaymath}
O_H = d (O_N).
\end{displaymath}
Let $P_k(z)= \tfrac{1}{k} Tr(\Phi^k)$, where the trace is in the adjoint representation. It is more convenient to work in the Katz-Morrison basis \cite{1992alg.geom..2002K}, related to the $P_k(z)$ by

\begin{align*}
\phi_2 = &\tfrac{1}{2^3\times 3\times 5}P_2\\
\phi_8=&-\tfrac{1}{2^7\times 3^2\times 5}\left(P_8-\tfrac{13}{2^8\times 3^3\times 5^4} P_2^4\right)\\
\phi_{12}=&\tfrac{1}{{2^6\times 3^4\times 5\times 7}}\left(P_{12}-\tfrac{3^2}{2^7\times 5^2} P_{8} P_{2}^2+\tfrac{7\times 101}{2^{15}\times 3^5\times 5^6} P_{2}^6\right)\\
\phi_{14}=&\tfrac{1}{2^6\times 3^2\times 5^2\times 7\times 11}\left(P_{14}-\tfrac{71}{2^4\times 3^2\times 5} P_{12} P_{2}+\tfrac{11\times 103}{2^{11}\times 3^3\times 5^3} P_{8} P_{2}^3-\tfrac{11\times 2531}{2^{19}\times 3^7\times 5^7} P_{2}^7\right)\\
\phi_{18}=&-\tfrac{1}{2^8\times 3^4\times 5^2\times 7\times 13}\left(P_{18}-\tfrac{1523}{2^6\times 3\times 5^3\times 11} P_{14} P_{2}^2+\tfrac{13\times 4451}{2^{10}\times 3^3\times 5^4\times 11} P_{12} P_{2}^3\right.\\
&\left.-\tfrac{13\times 331}{2^{11}\times 3^2\times 5^2} P_{8}^2 P_{2}-\tfrac{13\times 47\times 101}{2^{18}\times 3^5\times 5^5} P_{8} P_{2}^5+\tfrac{13\times 26399}{2^{27}\times 3^7\times 5^9} P_{2}^9\right)\\
\phi_{20} = & -\tfrac{1}{2^7\times 3^2\times 5^2\times 11\times 17\times 41}\left(P_{20} - \tfrac{236627}{2^5\times 3^3\times 5^2\times 7\times 13} P_{18}P_{2}+\tfrac{17\times 67\times 17389}{2^11\times 3^4\times 5^5\times 7\times 13} P_{14}P_{2}^3\right.\\
&\left.-\tfrac{17\times 127}{2^7\times  3\times 5\times 7}P_{12} P_{8}-\tfrac{17\times 191071}{2^11\times 3^6\times 5^6\times 7} P_{12} P_{2}^4+\tfrac{11\times 17\times 323371}{2^16\times 3^5\times 5^4\times 7} P_{8}^2 P_{2}^2\right.\\
&\left.+\tfrac{11\times 17\times 907\times 3301}{2^23\times 3^8\times 5^7\times 7} P_{8} P_{2}^6-\tfrac{11\times 13^2\times 17\times 60649}{2^32\times 3^11\times 5^10\times 7} P_{2}^{10}\right)\\
\phi_{24} =&\tfrac{1}{2^8\times 3^3\times 5\times 7\times 11\times 19\times 199}\left(P_{24}-\tfrac{7^2\times 95189}{2^7\times 3^2\times 5^4\times 17\times 41} P_{20} P_{2}^2+\tfrac{19\times 101\times 151\times 36821}{2^{12}\times 3^3\times 5^6\times 13\times 17\times 41}P_{18} P_{2}^3\right.\\
&\left.-\tfrac{19\times 31\times 751}{2^{10}\times 3^3\times 5^4} P_{14} P_{8} P_{2}-\tfrac{19\times 349\times 26280427}{2^{17}\times 3^6\times 5^9\times 13\times 41} P_{14} P_{2}^5-\tfrac{19\times 193}{2^6\times 3\times 5^2\times 7}P_{12}^2\right.\\
&\left.+\tfrac{19\times 269\times 870593}{2^{13}\times 3^5\times 5^5\times 7\times 41}P_{12} P_{8} P_{2}^2+\tfrac{19\times 199\times 379\times 602297}{2^{22}\times 3^8\times 5^{10}\times 41}P_{12} P_{2}^6-\tfrac{11\times 19\times 593}{2^14\times 3^2\times 5^3} P_{8}^3\right.\\
&\left.-\tfrac{11\times 19^2\times 891351347}{2^{23}\times 3^6\times 5^8\times 7\times 41} P_{8}^2 P_{2}^4-\tfrac{11\times 19\times 1489\times 14087599}{2^{29}\times 3^{10}\times 5^{11}\times 41} P_{8} P_{2}^8+\tfrac{19\times 331\times 49871\times 500881}{2^{39}\times 3^{13}\times 5^{15}\times 41} P_{2}^{12}\right)\\
\phi_{30}=&-\tfrac{1}{2^7\times 3^4\times 5^5\times 7\times 11\times 13\times 61}\left(P_{30}-\tfrac{13\times 389\times 54829}{2^13\times 3^2\times 5^3\times 11\times 19\times 199}  P_{24}P_{2}^3-\tfrac{7^3\times 13\times 1483}{2^11\times 3^2\times 5^2\times 11\times 41}P_{20} P_{8} P_{2}\right.\\
&\left. +\tfrac{7\times 13\times 9787\times 220623541}{2^20\times 3^5\times 5^7\times 11\times 19\times 41\times 199}P_{20} P_{2}^5 -\tfrac{2521}{2^8\times 3^2\times 7} P_{18}P_{12} +\tfrac{83\times 12666877}{2^15\times 3^5\times 5^2\times 7\times 11\times 41} P_{18}P_{8} P_{2}^2\right.\\
&\left. -\tfrac{1873\times 1118261\times 7442117}{2^25\times 3^8\times 5^9\times 11\times 41\times 199} P_{18}P_{2}^6 -\tfrac{17\times 1676569}{2^11\times 3^2\times 5^3\times 7\times 11^2}  P_{14}^2 P_{2}+\tfrac{19\times 9953023}{2^14\times 3^4\times 5^3\times 7\times 11^2} P_{14} P_{12} P_{2}^2\right.\\
&\left. -\tfrac{13\times 1223}{2^14\times 3^2\times 5}P_{14} P_{8}^2 +\tfrac{3442332938170993}{2^23\times 3^6\times 5^7\times 7\times 11\times 41\times 199} P_{14} P_{8} P_{2}^4+\tfrac{23\times 71065807756850923}{2^29\times 3^9\times 5^12\times 7\times 11\times 41\times 199} P_{14} P_{2}^8\right.\\
&\left. +\tfrac{13\times 89\times 7149761}{2^18\times 3^7\times 5^2\times 7\times 11^2\times 199} P_{12}^2 P_{2}^3 +\tfrac{13\times 1381\times 144773}{2^19\times 3\times 5^3\times 7\times 11\times 41} P_{12} P_{8}^2 P_{2} -\tfrac{13\times 31\times 90659\times 133346231}{2^23\times 3^7\times 5^8\times 7\times 11\times 41\times 199} P_{12} P_{8} P_{2}^5\right.\\
&\left. -\tfrac{13\times 9059\times 3294013\times 4114741}{2^34\times 3^12\times 5^13\times 7\times 11\times 41} P_{12} P_{2}^9 -\tfrac{13\times 1609\times 139987568111}{2^27\times 3^8\times 5^6\times 7\times 41\times 199} P_{8}^3 P_{2}^3 +\tfrac{13\times 17\times 2273\times 4259\times 325188503}{2^36\times 3^10\times 5^11\times 7\times 199} P_{8}^2 P_{2}^7\right.\\
&\left. +\tfrac{13\times 966205043352894287}{2^39\times 3^13\times 5^14\times 7\times 41\times 199} P_{8} P_{2}^{11} -\tfrac{29\times 71\times 10477\times 43777\times 41018108983}{2^52\times 3^16\times 5^18\times 7\times 11\times 41\times 199} P_{2}^{15}\right)
\end{align*}
The $\phi_k(z)$ have poles at the punctures of order at most $p_k$:

\begin{displaymath}
\phi_k(z) = \sum_{j=1}^{p_k} \frac{c^{(k)}_j}{z^j} + \text{regular}
\end{displaymath}
The $c^{(k)}_j$ are gauge-invariant parameters which contribute to the (graded) dimension(s) of the Coulomb branch. Were they all independent, the contribution would be $p_k$. Instead, the $c^{(k)}_j$ typically obey an elaborate set of polynomial relations. When the special piece of $O_N$ has more than one element, we have an additional quotient by a finite group (the ``Sommers-Achar group'') acting on the coefficients.

The detailed form of the constraints will be presented elsewhere \cite{Chacaltana:2018xxx}. For present purposes, it will suffice to list the \emph{net} contribution $n_k$ to the graded Coulomb branch dimension, which results after solving the constraints.

For each puncture, we also list its contribution to $n_h=4(5c-4a)$ and $n_v=4(2a-c)$.

\subsection{Regular Punctures}\label{regular_punctures}

Here, we list the pole structures, $\{p_2,p_8,p_{12},p_{14},p_{18},p_{20},p_{24},p_{30}\}$ of the 69 regular punctures of the $E_8$ theory, their graded Coulomb branch dimensions,\hfil\break $\{n_2,n_3,n_4,n_5,n_6,n_8,n_9,n_{10},n_{12},n_{14},n_{15},n_{18},n_{20},n_{24},n_{30}\}$, flavour symmetries and $(n_h,n_v)$.

{
\scriptsize
\setlength\LTleft{-.375in}
\renewcommand{\arraystretch}{2.25}

\begin{longtable}{|c|c|c|c|c|c|}
\hline
\mbox{\shortstack{Nahm\\B-C label}}&\mbox{\shortstack{Hitchin\\B-C label}}&Pole structure&Graded Coulomb Branch Dimensions&Flavour group&$(\delta n_h,\delta n_v)$\\
\hline 
\endhead
$0$&$E_8$&$\{1, 7, 11, 13, 17, 19, 23, 29\}$&$\{1, 0, 0, 0, 0, 7, 0, 0, 11, 13, 0, 17, 19, 23, 29\}$&${(E_8)}_{60}$&$(4960,4840)$\\
\hline
$A_1$&$E_8(a_1)$&$\{1, 7, 11, 13, 17, 19, 23, 28\}$&$\{1, 0, 0, 0, 0, 7, 0, 0, 11, 13, 0, 17, 19, 23, 28\}$&${(E_7)}_{48}$&$(4872,4781)$\\
\hline
$2A_1$&$E_8(a_2)$&$\{1, 7, 11, 13, 17, 19, 22, 28\}$&$\{1, 0, 0, 0, 0, 7, 0, 0, 11, 13, 0, 17, 19, 22, 28\}$&${SO(13)}_{40}$&$(4808,4734)$\\
\hline
$3A_1$&$(E_8(a_3),\mathbb{Z}_2)$&$\{1, 7, 11, 13, 17, 18, 22, 28\}$&$\{1, 0, 0, 0, 0, 7, 0, 0, 11, 13, 0, 17, 18, 22, 28\}$&${(F_4)}_{36}\times SU(2)_{31}$&$(4759,4695)$\\\hline
$A_2$&$E_8(a_3)$&$\{1, 7, 11, 13, 17, 18, 22, 28\}$&$\{1, 0, 0, 0, 0, 7, 0, 0, 11, 13, 1, 17, 18, 22, 27\}$&${(E_6)}_{36}$&$(4728,4665)$\\\hline
$4A_1$&$(E_8(a_4),\mathbb{Z}_2)$&$\{1, 7, 11, 13, 16, 18, 22, 28\}$&$\{1, 0, 0, 0, 0, 7, 0, 0, 11, 13, 0, 16, 18, 22, 28\}$&${Sp(4)}_{31}$&$(4716,4660)$\\\hline
$A_2+A_1$&$E_8(a_4)$&$\{1, 7, 11, 13, 16, 18, 22, 28\}$&$\{1, 0, 0, 0, 0, 7, 0, 0, 11, 13, 1, 16, 18, 22, 27\}$&${SU(6)}_{30}$&$(4682,4630)$\\\hline
$A_2+2A_1$&$E_8(b_4)$&$\{1, 7, 11, 13, 16, 18, 22, 27\}$&$\{1, 0, 0, 0, 0, 7, 0, 0, 11, 13, 0, 16, 18, 22, 27\}$&${SO(7)}_{28}\times {SU(2)}_{144}$&$(4648,4601)$\\\hline
$A_2+3A_1$&$(E_8(a_5),\mathbb{Z}_2)$&$\{1,7,11,12,16,18,22,27\}$&$\{1, 0, 0, 0, 0, 7, 0, 0, 11, 12, 0, 16, 18, 22, 27\}$&${(G_2)}_{48}\times {SU(2)}_{25}$&$(4617,4574)$\\\hline
$2A_2$&$E_8(a_5)$&$\{1, 7, 11, 12, 16, 18, 22, 27\}$&$\{1, 0, 0, 0, 0, 7, 0, 0, 12, 12, 0, 16, 18, 21, 27\}$&${(G_2)}^2_{24}$&$(4592,4550)$\\\hline
$A_3$&$E_7(a_1)$&$\{1, 7, 11, 13, 16, 18, 22, 27\}$&$\{1, 0, 0, 0, 0, 7, 0, 1, 11, 13, 0, 16, 18, 21, 26\}$&${SO(11)}_{28}$&$(4560,4514)$\\\hline
$2A_2+A_1$&$(E_8(b_5),S_3)$&$\{1, 7, 11, 12, 16, 18, 21, 27\}$&$\{1, 0, 0, 0, 0, 7, 0, 0, 11, 12, 0, 16, 18, 21, 27\}$&${(G_2)}_{24}\times {SU(2)}_{62}$&$(4566,4527)$\\\hline
$A_3+A_1$&$(E_8(b_5),\mathbb{Z}_2)$&$\{1, 7, 11, 12, 16, 18,21,27\}$&$\{1, 0, 0, 0, 0, 7, 0, 1, 11, 12, 0, 16, 18, 21, 26\}$&${SO(7)}_{24}\times {SU(2)}_{21}$&$(4525,4487)$\\\hline
$D_4(a_1)$&$E_8(b_5)$&$\{1, 7, 11, 12, 16, 18, 21, 27\}$&$\{1, 0, 0, 0, 0, 7, 0, 2, 11, 12, 0, 16, 17, 21, 26\}$&${SO(8)}_{24}$&$(4504,4467)$\\\hline
$2A_2+2A_1$&$(E_8(a_6),S_3)$&$\{1, 7, 10, 12, 16, 18, 21, 27\}$&$\{1, 0, 0, 0, 0, 7, 0, 0, 10, 12, 0, 16, 18, 21, 27\}$&${Sp(2)}_{62}$&$(4540,4504)$\\\hline
$A_3+2A_1$&$(E_8(a_6),\mathbb{Z}_2)$&$\{1, 7, 10, 12,16,18,21,27\}$&$\{1, 0, 0, 0, 0, 7, 0, 1, 10, 12, 0, 16, 18, 21, 26\}$&${Sp(2)}_{21}\times {SU(2)}_{40}$&$(4498,4464)$\\\hline
$D_4(a_1)+A_1$&$E_8(a_6)$&$\{1, 7, 10, 12, 16, 18, 21, 27\}$&$\{1, 0, 0, 0, 0, 7, 0, 2, 10, 12, 0, 16, 17, 21, 26\}$&${SU(2)}_{20}^3$&$(4476,4444)$\\\hline
$D_4$&$E_6$&$\{1, 7, 11, 12, 16, 17, 21, 26\}$&$\{1, 0, 0, 0, 1, 7, 0, 0, 11, 12, 0, 16, 16, 20, 24\}$&$(F_4)_{24}$&$(4272,4236)$\\\hline
$A_3+A_2$&$D_7(a_1)$&$\{1, 7, 10, 12, 16, 18, 21, 27\}$&$\{1, 0, 0, 0, 0, 7, 0, 1, 10, 12, 0, 16, 17, 21, 26\}$&${Sp(2)}_{20}\times U(1)_{48}$&$(4456,4425)$\\\hline
$A_4$&$E_7(a_3)$&$\{1, 7, 10, 12, 16, 18, 21, 27\}$&$\{1, 0, 0, 0, 0, 7, 1, 1, 10, 12, 1, 15, 17, 20, 25\}$&${SU(5)}_{20}$&$(4360,4330)$\\\hline
$A_3+A_2+A_1$&$(E_8(b_6),\mathbb{Z}_2)$&$\{1,7,10,12,16,17,21,26\}$&$\{1, 0, 0, 0, 0, 7, 0, 0, 10, 12, 0, 16, 17, 21, 26\}$&$SU(2)_{384}\times SU(2)_{19}$&$(4435,4406)$\\\hline
$D_4(a_1)+A_2$&$E_8(b_6)$&$\{1, 7, 10, 12, 16, 17, 21, 26\}$&$\{1, 0, 0, 0, 0, 7, 1, 0, 10, 12, 0, 15, 17, 21, 26\}$&$SU(3)_{96}$&$(4416,4388)$\\\hline
$A_4+A_1$&$E_6(a_1)+A_1$&$\{1, 7, 10, 12, 16, 17, 21, 26\}$&$\{1, 0, 0, 0, 0, 7, 1, 0, 10, 12, 1, 15, 17, 20, 25\}$&$SU(3)_{18}\times U(1)$&$(4337,4311)$\\\hline
$D_4+A_1$&$(E_6(a_1),\mathbb{Z}_2)$&$\{1, 7, 10, 12, 16,17,21,26\}$&$\{1, 0, 0, 0, 1, 7, 0, 0, 10, 12, 0, 16, 16, 20, 24\}$&$Sp(3)_{19}$&$(4241,4213)$\\\hline
$D_5(a_1)$&$E_6(a_1)$&$\{1, 7, 10, 12, 16, 17, 21, 26\}$&$\{1, 0, 0, 0, 1, 7, 1, 0, 10, 12, 0, 15, 16, 20, 24\}$&$SU(4)_{18}$&$(4220,4195)$\\\hline
$2A_3$&$(D_7(a_2),\mathbb{Z}_2)$&$\{1, 7, 10, 12, 15, 17, 20, 26\}$&$\{1, 0, 0, 0, 0, 7, 0, 0, 10, 12, 0, 15, 17, 20, 26\}$&$Sp(2)_{31}$&$(4350,4324)$\\\hline
$A_4+2A_1$&$D_7(a_2)$&$\{1, 7, 10, 12, 15, 17, 20, 26\}$&$\{1, 0, 0, 0, 0, 7, 0, 0, 10, 12, 1, 15, 17, 20, 25\}$&$SU(2)_{30}\times U(1)$&$(4318,4294)$\\\hline
$A_4+A_2$&$D_5+A_2$&$\{1, 7, 10, 12, 15, 17, 20, 25\}$&$\{1, 0, 0, 0, 0, 7, 0, 0, 10, 12, 0, 15, 17, 20, 25\}$&$SU(2)_{16}\times SU(2)_{200}$&$(4288,4265)$\\\hline
$D_5(a_1)+A_1$&$E_7(a_4)$&$\{1, 7, 10, 12, 15, 17, 20, 25\}$&$\{1, 0, 0, 0, 1, 7, 0, 0, 10, 12, 0, 15, 16, 20, 24\}$&$SU(2)_{16}\times SU(2)_{112}$&$(4200,4178)$\\\hline
$A_4+A_2+A_1$&$A_6+A_1$&$\{1, 6, 10, 12, 15, 17, 20, 25\}$&$\{1, 0, 0, 0, 0, 6, 0, 0, 10, 12, 0, 15, 17, 20, 25\}$&$SU(2)_{200}$&$(4272,4250)$\\\hline
$D_4+A_2$&$A_6$&$\{1, 6, 10, 12, 15, 17, 20, 25\}$&$\{1, 0, 0, 0, 1, 6, 0, 0, 10, 12, 0, 15, 16, 20, 24\}$&$SU(3)_{28}$&$(4184,4163)$\\\hline
$A_5$&$(D_6(a_1),\mathbb{Z}_2)$&$\{1, 7, 10, 12, 15, 17, 20, 25\}$&$\{1, 0, 0, 0, 1, 7, 0, 0, 11, 11, 0, 15, 16, 19, 24\}$&${(G_2)}_{16}\times {SU(2)}_{13}$&$(4149,4127)$\\\hline
$E_6(a_3)$&$D_6(a_1)$&$\{1, 7, 10, 12, 15, 17, 20, 25\}$&$\{1, 0, 0, 0, 2, 7, 0, 0, 10, 11, 0, 15, 16, 19, 24\}$&$(G_2)_{16}$&$(4136,4115)$\\\hline
$D_5$&$D_5$&$\{1, 7, 10, 12, 15, 17, 20, 25\}$&$\{1, 0, 0, 0, 1, 7, 0, 1, 10, 11, 0, 14, 15, 18, 22\}$&$SO(7)_{16}$&$(3904,3884)$\\\hline
$A_4+A_3$&$(E_8(a_7),S_5)$&$\{1, 6, 10, 11, 15, 16, 20, 25\}$&$\{1, 0, 0, 0, 0, 6, 0, 0, 10, 11, 0, 15, 16, 20, 25\}$&$SU(2)_{124}$&$(4204,4184)$\\\hline
$A_5+A_1$&$(E_8(a_7),S_3\times\mathbb{Z}_2)$&$\{1,6,10,11,15,16,20,25\}$&$\{1, 0, 0, 0, 2, 6, 0, 0, 10, 11, 0, 14, 16, 20, 24\}$&$SU(2)_{38}\times SU(2)_{13}$&$(4131,4112)$\\\hline
$D_5(a_1)+A_2$&$(E_8(a_7),S_4)$&$\{1, 6, 10, 11, 15, 16, 20, 25\}$&$\{1, 0, 0, 0, 1, 6, 0, 0, 10, 11, 0, 15, 16, 20, 24\}$&$SU(2)_{75}$&$(4155,4136)$\\\hline
$D_6(a_2)$&$(E_8(a_7),\mathbb{Z}_2\times\mathbb{Z}_2)$&$\{1,6,10,11,15,16,20,25\}$&$\{1, 0, 0, 0, 3, 6, 0, 0, 9, 11, 0, 15, 16, 19, 24\}$&$SU(2)_{13}^2$&$(4106,4088)$\\\hline
$E_6(a_3)+A_1$&$(E_8(a_7),S_3)$&$\{1, 6, 10, 11, 15, 16, 20, 25\}$&$\{1, 0, 0, 0, 2, 6, 0, 0, 10, 11, 0, 15, 16, 19, 24\}$&$SU(2)_{38}$&$(4118,4100)$\\\hline
$E_7(a_5)$&$(E_8(a_7),\mathbb{Z}_2)$&$\{1, 6, 10, 11,15,16,20,25\}$&$\{1, 0, 0, 0, 3, 6, 0, 0, 10, 11, 0, 14, 16, 19, 24\}$&$SU(2)_{13}$&$(4093,4076)$\\\hline
$E_8(a_7)$&$E_8(a_7)$&$\{1, 6, 10, 11, 15, 16, 20, 25\}$&$\{1, 0, 0, 0, 4, 6, 0, 0,  9, 11, 0, 14, 16, 19, 24\}$&$-$&$(4080,4064)$\\\hline
$A_6$&$D_4+A_2$&$\{1, 6, 10, 11, 15, 16, 20, 25\}$&$\{1, 0, 0, 0, 1, 6, 0, 1,  9, 11, 0, 14, 15, 18, 23\}$&$SU(2)_{12}\times SU(2)_{60}$&$(3920,3905)$\\\hline
$D_5+A_1$&$(E_6(a_3),\mathbb{Z}_2)$&$\{1, 6, 10, 11, 15,16,20,25\}$&$\{1, 0, 0, 0, 1, 6, 0, 1, 10, 11, 0, 14, 15, 18, 22\}$&$SU(2)_{13}\times SU(2)_{24}$&$(3885,3869)$\\\hline
$D_6(a_1)$&$E_6(a_3)$&$\{1, 6, 10, 11, 15, 16, 20, 25\}$&$\{1, 0, 0, 0, 2, 6, 0, 1, 9, 11, 0, 14, 15, 18, 22\}$&$SU(2)_{12}^2$&$(3872,3857)$\\\hline
$A_6+A_1$&$A_4+A_2+A_1$&$\{1, 6,  9, 11, 14, 16, 19, 24\}$&$\{1, 0, 0, 0, 0, 6, 0, 1, 9, 11, 0, 14, 15, 18, 23\}$&$SU(2)_{60}$&$(3908,3894)$\\\hline
$E_7(a_4)$&$D_5(a_1)+A_1$&$\{1, 6, 10, 11, 15, 16, 20, 25\}$&$\{1, 0, 0, 0, 1, 6, 0, 1, 9, 11, 0, 14, 15, 18, 22\}$&$SU(2)_{12}$&$(3860,3846)$\\\hline
$E_6(a_1)$&$D_5(a_1)$&$\{1, 6, 10, 11, 15, 16, 20, 25\}$&$\{1, 0, 0, 1, 1, 6, 1, 0, 9, 10, 1, 13, 14, 17, 21\}$&$SU(3)_{12}$&$(3688,3675)$\\\hline
$D_5+A_2$&$A_4+A_2$&$\{1, 6,  9, 11, 14, 16, 19, 24\}$&$\{1, 0, 0, 0, 0, 6, 0, 1, 9, 11, 0, 14, 15, 18, 22\}$&$U(1)$&$(3848,3835)$\\\hline
$E_6$&$D_4$&$\{1, 6, 10, 11, 15, 16, 20, 25\}$&$\{1, 0, 0, 0, 2, 6, 0, 0, 9,  9, 0, 12, 12, 15, 18\}$&${(G_2)}_{12}$&$(3232,3220)$\\\hline
$D_7(a_2)$&$A_4+2A_1$&$\{1, 6, 9, 11, 14, 16, 19, 24\}$&$\{1, 0, 0, 1, 0, 6, 1, 0, 9, 10, 0, 13, 15, 18, 22\}$&$U(1)$&$(3792,3780)$\\\hline
$E_6(a_1)+A_1$&$A_4+A_1$&$\{1, 6, 9, 11, 14, 16, 19, 24\}$&$\{1, 0, 0, 1, 0, 6, 1, 0, 9, 10, 1, 13, 14, 17, 21\}$&$U(1)$&$(3675,3664)$\\\hline
$D_6$&$(A_4,\mathbb{Z}_2)$&$\{1, 6, 9, 11, 14, 16, 19, 24\}$&$\{1, 0, 1, 0, 0, 6, 0, 2, 8, 10, 0, 12, 14, 16, 20\}$&${Sp(2)}_{11}$&$(3502,3490)$\\\hline
$E_7(a_3)$&$A_4$&$\{1, 6, 9, 11, 14, 16, 19, 24\}$&$\{1, 0, 1, 1, 0, 6, 0, 1, 8, 10, 0, 12, 14, 16, 20\}$&${SU(2)}_{10}$&$(3490,3480)$\\\hline
$A_7$&$(D_4(a_1)+A_2,\mathbb{Z}_2)$&$\{1, 6, 9, 10, 13,14,17,22\}$&$\{1, 0, 0, 0, 0, 6, 0, 0, 9, 10, 0, 13, 14, 17, 22\}$&${SU(2)}_{31}$&$(3679,3668)$\\\hline
$E_8(b_6)$&$D_4(a_1)+A_2$&$\{1, 6, 9, 10, 13, 14, 17, 22\}$&$\{1, 0, 0, 0, 0, 6, 0, 0, 9, 10, 1, 13, 14, 17, 21\}$&$-$&$(3648,3638)$\\\hline
$D_7(a_1)$&$A_3+A_2$&$\{1, 6, 9, 10, 13, 14, 17, 21\}$&$\{1, 0, 1, 0, 0, 6, 0, 1, 8, 10, 0, 12, 14, 16, 20\}$&$U(1)$&$(3480,3471)$\\\hline
$E_8(a_6)$&$D_4(a_1)+A_1$&$\{1, 6, 9, 10, 13, 14, 17, 21\}$&$\{1, 0, 2, 0, 0, 5, 0, 2, 8, 9, 0, 12, 13, 16, 20\}$&$-$&$(3424,3416)$\\\hline
$E_6+A_1$&$(D_4(a_1),S_3)$&$\{1, 6, 9, 10, 13, 14, 17, 21\}$&$\{1, 0, 0, 0, 1, 6, 0, 0, 9, 9, 0, 12, 12, 15, 18\}$&${SU(2)}_{26}$&$(3218,3209)$\\\hline
$E_7(a_2)$&$(D_4(a_1),\mathbb{Z}_2)$&$\{1, 6, 9, 10,13,14,17,21\}$&$\{1, 0, 1, 0, 1, 6, 0, 0, 8, 9, 0, 12, 12, 15, 18\}$&${SU(2)}_{9}$&$(3201,3193)$\\\hline
$E_8(b_5)$&$D_4(a_1)$&$\{1, 6, 9, 10, 13, 14, 17, 21\}$&$\{1, 0, 2, 0, 1, 5, 0, 0, 8, 9, 0, 12, 12, 15, 18\}$&$-$&$(3192,3185)$\\\hline
$D_7$&$(2A_2,\mathbb{Z}_2)$&$\{1, 5, 8,  9, 12, 13, 16, 20\}$&$\{1, 0, 0, 0, 1, 5, 0, 0, 8, 8, 0, 11, 12, 14, 18\}$&$SU(2)_{13}$&$(3069,3062)$\\\hline
$E_8(a_5)$&$2A_2$&$\{1, 5, 8,  9, 12, 13, 16, 20\}$&$\{1, 0, 0, 0, 2, 5, 0, 0, 7, 8, 0, 11, 12, 14, 18\}$&$-$&$(3056,3050)$\\\hline
$E_7(a_1)$&$A_3$&$\{1, 6, 9, 10, 13, 14, 17, 21\}$&$\{1, 0, 1, 0, 1, 5, 0, 1, 7, 8, 0, 10, 11, 13, 16\}$&${SU(2)}_8$&$(2832,2826)$\\\hline
$E_8(b_4)$&$A_2+2A_1$&$\{1, 5, 8, 9, 12, 13, 16, 20\}$&$\{1, 0, 0, 0, 1, 5, 0, 1, 7, 8, 0, 10, 11, 13, 16\}$&$-$&$(2824,2819)$\\\hline
$E_8(a_4)$&$A_2+A_1$&$\{1, 5, 8, 9, 12, 13, 16, 20\}$&$\{1, 1, 0, 1, 0, 4, 1, 0, 6, 7, 1, 9, 10, 12, 15\}$&$-$&$(2608,2604)$\\\hline
$E_7$&$(A_2,\mathbb{Z}_2)$&$\{1, 5, 8, 9, 12, 13, 16, 20\}$&$\{1, 0, 0, 0, 2, 4, 0, 0, 6, 6, 0, 8, 8, 10, 12\}$&${SU(2)}_7$&$(2159,2155)$\\\hline
$E_8(a_3)$&$A_2$&$\{1, 5, 8, 9, 12, 13, 16, 20\}$&$\{1, 1, 0, 0, 1, 4, 0, 0, 6, 6, 0, 8, 8, 10, 12\}$&$-$&$(2152,2149)$\\\hline
$E_8(a_2)$&$2A_1$&$\{1, 4, 6, 7, 9, 10, 12, 15\}$&$\{1, 0, 1, 0, 0, 3, 0, 1, 4, 5, 0, 6, 7, 8, 10\}$&$-$&$(1752,1750)$\\\hline
$E_8(a_1)$&$A_1$&$\{1, 3, 4, 5, 6, 7, 8, 10\}$&$\{1, 0, 0, 0, 1, 2, 0, 0, 3, 3, 0, 4, 4, 5, 6\}$&$-$&$(1080,1079)$\\\hline
\end{longtable}
}

\subsection{Irregular Punctures and Irregular Fixtures}\label{irregular_punctures_and_irregular_fixtures}

An irregular puncture, in the sense of \cite{Chacaltana:2012zy}, is a pair: $(\mathcal{O}, H_{k'})$, where $\mathcal{O}$ is a regular puncture, $H_{k'}$ is a simple subgroup of the global symmetry group $F(\mathcal{O})$, and $0\leq k' = 4\kappa_H - k$ where $\kappa_H$ is the dual Coxeter number.

For the $E_8$ theory, there are 30 irregular punctures, all but 3 of which actually appear in the irregular fixtures below. The remaining ones, $(A_1, {Spin(12)}_8)$, $(A_1, {Spin(11)}_4)$, and $(A_1, {Spin(10)}_0)$, appear to play no role in the theory.

To each irregular puncture, $(\mathcal{O}, H_{k'})$, we assign

\begin{displaymath}
\begin{aligned}
n_h &= 9920 -n_h(\mathcal{O})\\
n_v &= 9928 -n_v(\mathcal{O})- \dim(H).\\
\end{aligned}
\end{displaymath}
An irregular fixture

\begin{displaymath}
 \includegraphics[width=93pt]{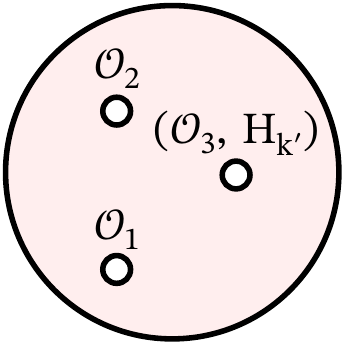}
\end{displaymath}
consists of

\begin{itemize}%
\item a pair of regular punctures, $\mathcal{O}_1$ and $\mathcal{O}_2$ for which\begin{displaymath}
 \includegraphics[width=93pt]{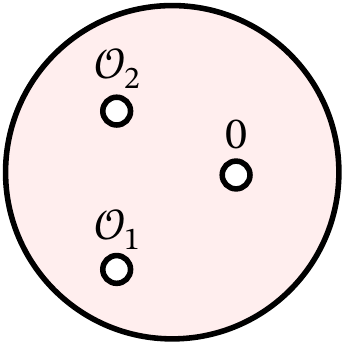}
\end{displaymath}
would be \emph{bad}.

\item an SCFT, $X$, (possibly free hypers, an isolated interacting SCFT, or a mixture) whose flavour symmetry\begin{displaymath}
F(X)\times \text{cent}(H_k\subset F(\mathcal{O}_3))\supset F(\mathcal{O}_1)\times F(\mathcal{O}_2)\times H_{k'}
\end{displaymath}
and whose central charges satisfy

\begin{displaymath}
\begin{aligned}
n_h(X)&= n_h(\mathcal{O}_1)+n_h(\mathcal{O}_2)-n_h(\mathcal{O}_3)\\
n_v(X)&= n_v(\mathcal{O}_1)+n_v(\mathcal{O}_2)-n_v(\mathcal{O}_3)-\dim(H).
\end{aligned}
\end{displaymath}

\end{itemize}
In fact, we have something stronger than the relation on $n_v$. A similar relation must hold on the graded Coulomb dimensions\footnote{Since $n_v = \sum_k (2k-1) n_k$, \eqref{CBdimsRel} implies the relation on $n_v$.} 

\begin{equation}
n_k(X) =
\begin{cases}
n_k(\mathcal{O}_1)+n_k(\mathcal{O}_2)-n_k(\mathcal{O}_3)-1& (k-1)\, \text{an exponent of}\, \mathfrak{h}\\
n_k(\mathcal{O}_1)+n_k(\mathcal{O}_2)-n_k(\mathcal{O}_3)&\text{otherwise}.
\end{cases}
\label{CBdimsRel}\end{equation}
Going through the list of pairs $\mathcal{O}_1,\mathcal{O}_2$, we readily identify the 50 irregular fixtures in the $E_8$ theory. In fact, to determine $(\mathcal{O}_3,H_{k'})$, it suffices to know that the relation on $n_v$ (rather than the more refined relations on the $n_k$) are satisfied. Given the $n_k(X)$ (in particular, for a free-field fixture, all the $n_k=0$), we thus obtain predictions for the $n_k(\mathcal{O}_i)$. These predictions are of assistance in computing the constraints on the pole-coefficients at the punctures -- by far the most tedious part of the computation. Since we won't compute any Seiberg-Witten solutions here, we will leave the detailed form of the constraints for a separate publication \cite{Chacaltana:2018xxx}, and report only the graded Coulomb branch dimensions here.

\subsubsection{Free-field Fixtures}\label{freefield_fixtures}

While there is no regular fixture consisting of free hypermultiplets (there are no Lagrangian field theories arising as the compactification of the $E_8$ theory), there are 29 fixtures, consisting of two regular punctures and an irregular puncture, which correspond to free hypermultiplets.

{
\renewcommand{\arraystretch}{1.75}
\begin{longtable}{|c|c|c|c|}
\hline
\#&Fixture&Number of hypers&Representation\\
\hline
\endhead
1&$\begin{matrix}E_8(a_1)\\ E_8(a_1)\end{matrix}\quad (E_7,{SU(2)}_1)$&$1$&$\tfrac{1}{2}(2)$\\
\hline
2&$\begin{matrix}E_8(a_1)\\ E_8(a_2)\end{matrix}\quad (E_7(a_1),{SU(2)}_0)$&$0$&empty\\
\hline
3&$\begin{matrix}E_8(a_1)\\ E_8(a_3)\end{matrix}\quad (E_6,{SU(3)}_0)$&$0$&empty\\
\hline
4&$\begin{matrix}E_8(a_1)\\ E_7\end{matrix}\quad (E_6,{(G_2)}_4)$&$7$&$1(7)$\\
\hline
5&$\begin{matrix}E_8(a_1)\\ E_8(a_4)\end{matrix}\quad (E_6(a_1),{SU(3)}_0)$&$0$&empty\\
\hline
6&$\begin{matrix}E_8(a_1)\\ E_8(b_4)\end{matrix}\quad (D_5,{(G_2)}_0)$&$0$&empty\\
\hline
7&$\begin{matrix}E_8(a_1)\\ E_7(a_1)\end{matrix}\quad (D_5,{Spin(7)}_4)$&$8$&$1(8)$\\
\hline
8&$\begin{matrix}E_8(a_1)\\ E_8(a_5)\end{matrix}\quad (E_6(a_3),{(G_2)}_0)$&$0$&empty\\
\hline
9&$\begin{matrix}E_8(a_1)\\ D_7\end{matrix}\quad (A_5,{(G_2)}_0)$&$0$&empty\\
\hline
10&$\begin{matrix}E_8(a_1)\\ E_8(b_5)\end{matrix}\quad (D_4,{Spin(8)}_0)$&$0$&empty\\
\hline
11&$\begin{matrix}E_8(a_1)\\ E_7(a_2)\end{matrix}\quad (D_4,{Spin(9)}_4)$&$9$&$1(9)$\\
\hline
12&$\begin{matrix}E_8(a_1)\\ E_6+A_1\end{matrix}\quad (D_4,{(F_4)}_{12})$&$26$&$1(26)$\\
\hline
13&$\begin{matrix}E_8(a_1)\\ E_8(a_6)\end{matrix}\quad (D_4(a_1),{Spin(8)}_0)$&$0$&empty\\
\hline
14&$\begin{matrix}E_8(a_1)\\ D_7(a_1)\end{matrix}\quad (A_3,{Spin(9)}_0)$&$0$&empty\\
\hline
15&$\begin{matrix}E_8(a_1)\\ E_8(b_6)\end{matrix}\quad (A_2,{(F_4)}_0)$&$0$&empty\\
\hline
16&$\begin{matrix}E_8(a_1)\\ A_7\end{matrix}\quad (3A_1,{(F_4)}_0)$&$0$&empty\\
\hline
17&$\begin{matrix}E_8(a_1)\\ E_7(a_3)\end{matrix}\quad (A_3,{Spin(10)}_4)$&$10$&$1(10)$\\
\hline
18&$\begin{matrix}E_8(a_1)\\ D_6\end{matrix}\quad (A_3,{Spin(11)}_8)$&$22$&$2(11)$\\
\hline
19&$\begin{matrix}E_8(a_1)\\ E_6(a_1)+A_1\end{matrix}\quad (A_2,{(E_6)}_{12})$&$27$&$1(27)$\\
\hline
20&$\begin{matrix}E_8(a_1)\\ D_7(a_2)\end{matrix}\quad (A_1,{(E_6)}_0)$&$0$&empty\\
\hline
21&$\begin{matrix}E_8(a_1)\\ D_5+A_2\end{matrix}\quad (A_1,{(E_7)}_{24})$&$56$&$1(56)$\\
\hline
22&$\begin{matrix}E_8(a_1)\\ A_6+A_1\end{matrix}\quad (0,{(E_7)}_{12})$&$28$&$\tfrac{1}{2}(56)$\\
\hline
23&$\begin{matrix}E_8(a_2)\\ E_8(a_2)\end{matrix}\quad (D_6,{Sp(2)}_{1})$&$2$&$\tfrac{1}{2}(4)$\\
\hline
24&$\begin{matrix}E_8(a_2)\\ E_8(a_3)\end{matrix}\quad (D_5,{SU(4)}_{0})$&$0$&empty\\
\hline
25&$\begin{matrix}E_8(a_2)\\ E_7\end{matrix}\quad (D_5,{Spin(7)}_{4})$&$7$&$1(7)$\\
\hline
26&$\begin{matrix}E_8(a_2)\\ E_8(a_4)\end{matrix}\quad (A_4,{SU(5)}_{0})$&$0$&empty\\
\hline
27&$\begin{matrix}E_8(a_2)\\ E_8(b_4)\end{matrix}\quad (A_3,{Spin(11)}_{8})$&$16$&$\tfrac{1}{2}(32)$\\
\hline
28&$\begin{matrix}E_8(a_2)\\ E_8(a_5)\end{matrix}\quad (2A_1,{Spin(12)}_{0})$&$0$&empty\\
\hline
29&$\begin{matrix}E_8(a_2)\\ D_7\end{matrix}\quad (2A_1,{Spin(13)}_{4})$&$13$&$1(13)$\\
\hline
\end{longtable}
}

\subsubsection{Interacting Fixtures with one Irregular Puncture}\label{interacting_fixtures_with_one_irregular_puncture}

There are 19 fixtures, consisting of two regular punctures and an irregular puncture, which correspond to interacting SCFTs. Most are (combinations of) Minahan-Nemeschansky theories; but the ${(F_4)}_{12}\times {SU(2)}_{7}^2$ SCFT (which first appeared in the $D_4$ theory \cite{Chacaltana:2011ze}) and the $(E_7)_{24}$ SCFT, the $(E_7)_{24}\times {SU(2)}_{7}$ SCFT, the $(E_7)_{24}\times {SU(2)}_{8}$ SCFT, the $(E_7)_{24}\times {SU(2)}_{8}\times {SU(2)}_{7}$ SCFT, the ${(E_7)}_{24}\times {Spin(7)}_{16}$ SCFT (all of which first appeared in the $E_6$ theory \cite{Chacaltana:2014jba}) and the ${(E_7)}_{24}\times {(G_2)}_{12}$ (which \href{https://golem.ph.utexas.edu/class-S/E7/fixtures/select?puncture1[id]=1&puncture2[id]=23&puncture3[id]=43}{appeared} in the $E_7$ theory \cite{Chacaltana:2017boe}) also make appearances.

Again, the graded Coulomb branch dimensions are $(n_2,n_3,n_4,n_6,n_8,n_{12},n_{14},n_{18},n_{20},n_{24},n_{30})$.

{
\footnotesize
\renewcommand{\arraystretch}{2.25}

\begin{longtable}{|c|l|c|c|l|}
\hline
\#&Fixture&Graded Coulomb branch dims&$(n_h,n_v)$&Theory\\
\hline 
\endhead
1&$\begin{matrix}E_8(a_1)\\ E_6\end{matrix}\quad (D_4,{(F_4)}_{12})$&$(0,0,0,1,0,0,0,0,0,0,0)$&$(40,11)$&${(E_8)}_{12}\, \text{SCFT}$\\ \hline 
2&$\begin{matrix}E_8(a_1)\\ E_6(a_1)\end{matrix}\quad (A_2,{(E_6)}_{12})$&$(0,0,0,1,0,0,0,0,0,0,0)$&$(40,11)$&${(E_8)}_{12}\, \text{SCFT}$\\ \hline 
3&$\begin{matrix}E_8(a_1)\\ D_6(a_1)\end{matrix}\quad (A_1,{(E_7)}_{24})$&$(0,0,0,2,0,0,0,0,0,0,0)$&$(80,22)$&${[{(E_8)}_{12}\, \text{SCFT}]}^2$\\ \hline 
4&$\begin{matrix}E_8(a_1)\\ D_5+A_1\end{matrix}\quad (A_1,{(E_7)}_{24})$&$(0,0,0,1,0,1,0,0,0,0,0)$&$(93,34)$&${(E_8)}_{24}\times {SU(2)}_{13}\, \text{SCFT}$\\ \hline 
5&$\begin{matrix}E_8(a_1)\\ A_6\end{matrix}\quad (0,{(E_7)}_{12})$&$(0,0,0,1,0,0,0,0,0,0,0)$&$(40,11)$&${(E_8)}_{12}\, \text{SCFT}$\\ \hline 
6&$\begin{matrix}E_8(a_1)\\ D_5\end{matrix}\quad (A_1,{(E_7)}_{24})$&$(0,0,0,1,1,1,0,0,0,0,0)$&$(112,49)$&${(E_7)}_{24}\times {Spin(7)}_{16}\, \text{SCFT}$\\ \hline 
7&$\begin{matrix}E_8(a_2)\\ E_7(a_1)\end{matrix}\quad (A_3,{Spin(11)}_{8})$&$(0,0,1,0,0,0,0,0,0,0,0)$&$(24,7)$&${(E_7)}_{8}\, \text{SCFT}$\\ \hline 
8&$\begin{matrix}E_8(a_2)\\ E_8(b_5)\end{matrix}\quad (A_1,{(E_7)}_{24})$&$(0,0,3,0,0,0,0,0,0,0,0)$&$(72,21)$&${[{(E_7)}_{8}\, \text{SCFT}]}^3$\\ \hline 
9&$\begin{matrix}E_8(a_2)\\ E_7(a_2)\end{matrix}\quad (A_1,{(E_7)}_{24})$&$(0,0,2,0,1,0,0,0,0,0,0)$&$(81,29)$&$\begin{gathered}
[{(E_7)}_{8}\, \text{SCFT}]\\ \times\\ [{(E_7)}_{16}\times {SU(2)}_{9}\, \text{SCFT}]\end{gathered}$\\ \hline 
10&$\begin{matrix}E_8(a_2)\\ E_6+A_1\end{matrix}\quad (A_1,{(E_7)}_{24})$&$(0,0,1,0,1,1,0,0,0,0,0)$&$(98,45)$&${(E_7)}_{24}\times {SU(2)}_{26}\, \text{SCFT}$\\ \hline 
11&$\begin{matrix}E_8(a_2)\\ E_6\end{matrix}\quad (A_1,{(E_7)}_{24})$&$(0,0,1,1,1,1,0,0,0,0,0)$&$(112,56)$&${(E_7)}_{24}\times {(G_2)}_{12}\, \text{SCFT}$\\ \hline 
12&$\begin{matrix}E_8(a_3)\\ E_8(a_3)\end{matrix}\quad (D_4,{(F_4)}_{12})$&$(0,2,0,0,0,0,0,0,0,0,0)$&$(32,10)$&${[{(E_6)}_{6}\, \text{SCFT}]}^2$\\ \hline 
13&$\begin{matrix}E_8(a_3)\\ E_7\end{matrix}\quad (D_4,{(F_4)}_{12})$&$(0,1,0,1,0,0,0,0,0,0,0)$&$(39,16)$&${(E_6)}_{12}\times {SU(2)}_7\, \text{SCFT}$\\ \hline 
14&$\begin{matrix}E_8(a_3)\\ E_8(a_4)\end{matrix}\quad (A_2,{(E_6)}_{12})$&$(0,2,0,0,0,0,0,0,0,0,0)$&$(32,10)$&${[{(E_6)}_{6}\, \text{SCFT}]}^2$\\ \hline 
15&$\begin{matrix}E_8(a_3)\\ E_8(b_4)\end{matrix}\quad (A_1,{(E_7)}_{24})$&$(0,1,0,1,1,1,0,0,0,0,0)$&$(104,54)$&${(E_7)}_{24}\, \text{SCFT}$\\ \hline 
16&$\begin{matrix}E_8(a_3)\\ E_7(a_1)\end{matrix}\quad (A_1,{(E_7)}_{24})$&$(0,1,1,1,1,1,0,0,0,0,0)$&$(112,61)$&${(E_7)}_{24}\times {SU(2)}_{8}\, \text{SCFT}$\\ \hline 
17&$\begin{matrix}E_7\\ E_7\end{matrix}\quad (D_4,{(F_4)}_{12})$&$(0,0,0,2,0,0,0,0,0,0,0)$&$(46,22)$&${(F_4)}_{12}\times {SU(2)}_{7}^2\, \text{SCFT}$\\ \hline 
18&$\begin{matrix}E_7\\ E_8(a_4)\end{matrix}\quad (A_2,{(E_6)}_{12})$&$(0,1,0,1,0,0,0,0,0,0,0)$&$(39,16)$&${(E_6)}_{12}\times {SU(2)}_{7}\, \text{SCFT}$\\ \hline 
19&$\begin{matrix}E_7\\ E_8(b_4)\end{matrix}\quad (A_1,{(E_7)}_{24})$&$(0,0,0,2,1,1,0,0,0,0,0)$&$(111,60)$&${(E_7)}_{24}\times {SU(2)}_{7}\, \text{SCFT}$\\ \hline 
20&$\begin{matrix}E_7\\ E_7(a_1)\end{matrix}\quad (A_1,{(E_7)}_{24})$&$(0,0,1,2,1,1,0,0,0,0,0)$&$(119,67)$&$\begin{gathered}{(E_7)}_{24}\times {SU(2)}_{8}\\ \times {SU(2)}_{7}\, \text{SCFT}\end{gathered}$\\ \hline 
\end{longtable}
}

\subsubsection{Mixed Fixtures with one Irregular Puncture}\label{mixed_fixtures_with_one_irregular_puncture}

There is one mixed fixture with two regular and one irregular puncture. The value of $n_h$, listed, is the one associated to the SCFT, \emph{after} subtracting the contribution of the free hypermultiplets.

{
\renewcommand{\arraystretch}{1.75}
\begin{longtable}{|c|l|c|c|c|}
\hline
\#&Fixture&Graded Coulomb Branch dims&$(n_h,n_v)$&Theory\\
\hline
\endhead
1&$\begin{matrix}E_8(a_1)\\ E_7(a_4)\end{matrix}\quad (A_1,{(E_7)}_{24})$&$(0,0,0,1,0,0,0,0,0,0,0)$&$(40,11)$&${(E_8)}_{12}\, \text{SCFT}$ + $\tfrac{1}{2}(56)$\\
\hline
\end{longtable}
}

\subsection{Fixtures}\label{fixtures}

The basic building block of the theories of class-S is the compactification of the (2,0) theory on a 3-punctured sphere (``fixture''). Given the 69 punctures of the $E_8$ theory, there are 57,155 possible fixtures. Of these, 7,319 are bad (do not give rise to a well-defined 4D SCFT). These are ``replaced'' by the 50 irregular fixtures listed above. The remaining 49,836 fixtures are too numerous to list here, but they can be explored in our \href{https://golem.ph.utexas.edu/class-S/E8/fixtures/}{online application}.

Our main task is to determine which of the 49,836 are mixed fixtures (containing free hypermultiplets + an interacting isolated SCFT) and which are isolated interacting SCFTs with an enhanced global symmetry

\begin{displaymath}
\begin{aligned}
   F &\supset F_{\text{manifest}}\\
   F_{\text{manifest}} &= F(\mathcal{O}_1)\times F(\mathcal{O}_2)\times F(\mathcal{O}_3).
\end{aligned}
\end{displaymath}
To this end, we compute the (refined) Hall-Littlewood index up to order $\tau^2$. The $O(\tau)$ contribution counts the number of free hypermultiplets. Removing the free hypermultiplets, the $O(\tau^2)$ contribution is the character $\chi(\text{adj}(F))$ decomposed as a representation of $F_{\text{manifest}}$. \cite{Gadde:2011uv, Gaiotto:2012uq}

We find 149 mixed fixtures and an additional 775 fixtures with enhanced global symmetries. The remaining 48,912 fixtures have global symmetry $F_{\text{manifest}}$.

\subsection{Computing the Hall-Littlewood Superconformal Index}\label{computing_the_halllittlewood_superconformal_index}

To classify the fixtures, as described in \S\ref{fixtures}, we need to calculate their superconformal index \cite{Kinney:2005ej,Gadde:2009kb,Gadde:2011ik,Gadde:2011uv,Lemos:2012ph}. Up to second order, the superconformal index of a fixture of type $G$ simplifies to:

\begin{equation}
I=1+\chi_{F_{manifest}}^{adj}\tau^2+\left.\left({\displaystyle{\sum_\lambda}'}\frac{\prod_{i=1}^3\chi^\lambda_i(a_i|\tau)}{\chi^\lambda_0}\right)\right\vert_{\tau^2}
\label{IndexSimp}\end{equation}
where $\chi^\lambda(a_i|\tau)$ is the Weyl character coming from puncture $i$ associated with the irreducible representation of $G$ whose highest weight is $\lambda$. $\tau$ is the fugacity of $SU(2)$ symmetry while $a_i$ are the fugacities of global symmetry $f$ of the puncture. The last term is an infinite sum over all nontrivial irreducible representation of $G$. The detailed form of $\chi$ depends on the branching rule of each puncture. However, due to the symmetric form of the character, it is easy to see that $\chi_i$ can always be written as

\begin{displaymath}
\frac{1}{\tau^{w_i}}(F_i(a_i)+O(\tau))
\end{displaymath}
where $w$ is the highest weight of the $SU(2)_R$ representation appearing in the branching rule of each puncture. $F(a_i)$ is a function of the fugacities of global symmetry. So we have

\begin{displaymath}
\left(\sum_\lambda\frac{\prod_{i=1}^3\chi^\lambda_i(a_i|\tau)}{\chi^\lambda_0}\right)=\sum_\lambda\tau^{w_0^\lambda-w_1^\lambda-w_2^\lambda-w_3^\lambda}(F_1^\lambda(a_1)F_2^\lambda(a_2)F_3^\lambda(a_3)/F_0^\lambda+O(\tau))
\end{displaymath}
If we can compute $W^\lambda\equiv w_0^\lambda-w_1^\lambda-w_2^\lambda-w_3^\lambda$ for a representation $\lambda$, we will know whether we need to include it in the sum at order $\tau^2$. Clearly we need to include $\lambda$ in the sum if and only if $W^\lambda\leq 2$.

Now observe that the projection matrix $p$ for each puncture projects the weights of an irreducible representation of $G$ onto the weights of a subgroup of $G$. For $SU(2)$, the weights are all just numbers and the highest weight is nothing but the largest of them. Let $V$ be the weight space of irreducible representation with highest weight $\lambda$, then:

\begin{displaymath}
w_i^\lambda=\text{max}_{v\in V}(p_i\cdot v)
\end{displaymath}
Because every $v$ can be represented as $\lambda$ minus a linear combination of simple positive roots $r$ of $G$. If we can choose $p$ such that $p\cdot r$ is nonnegative for all simple positive roots\footnote{A choice which does the trick is to take $p$ to be the weighted Dynkin diagram for the puncture times the inverse of the Cartan matrix.} , we will be able to compute $w$ and $W$ easily:

\begin{displaymath}
\begin{aligned}
w_i^\lambda&=p_i\cdot\lambda\\
W^\lambda&=(p_0-p_1-p_2-p_3)\cdot\lambda
\end{aligned}
\end{displaymath}
Let $P=p_0-p_1-p_2-p_3$. The vector is determined only by the punctures and can be easily computed for all fixtures. We have four cases for $P$:

\begin{enumerate}%
\item If any entry of $P$ is non-positive, then the SCI diverges and the fixture is bad. For example, consider a representation with $\lambda=(0\dots 0 10\dots 0)$ where the position of ``$1$" matches the position of the non-positive entry of $P$, this will lead to a term with non-positive power in $\tau$.
\item If all entries of $P$ are greater than $2$, then the fixture is an SCFT with no symmetry enhancements or free hypermultiplets.
\item If all entries of $P$ are positive and at least one of them is $1$ (and some other elements might be $2$), the SCFT will have free hypermultiplets and possibly will have enhanced symmetry. To compute this, we need to consider the $\lambda=(0\dots 010\dots 0)$, $(0\dots 020\dots 0)$ and $(0\dots 010\dots 010\dots 0)$ where the position of the nonzero elements matches the $1$'s in $P$. We also need to consider $\lambda=(0\dots 010\dots 0)$ where the position of $1$ matches the $2$'s in $P$.
\item If all elements of $P$ are greater than $1$ and at least one of them is $2$, the SCFT will have enhanced symmetry and no free hypermultiplets. We need to consider $\lambda=(0\dots 010\dots 0)$ where the position of $1$ matches the $2$s in $P$.

\end{enumerate}
Note that all representations we need to consider are of the form $(0\dots 010\dots 0)$, $(0\dots 020\dots 0)$ or $(0\dots 010\dots 010\dots 0)$. Therefore, despite the appearance, \eqref{IndexSimp} is actually a finite sum over $\lambda$.

As an example, consider a fixture of $E_7$ theory: $\begin{matrix} \includegraphics[width=76pt]{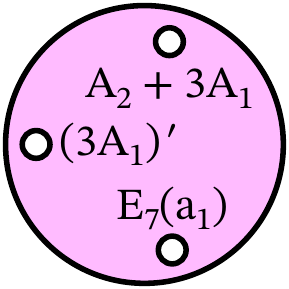}\end{matrix}$.

\begin{displaymath}
P_{{(3A_1)}', A_2+3A_1, E_7(a_1)}=p_{E_7}-p_{{(3A_1)}'}-p_{A_2+3A_1}-p_{E_7(a_1)}=(1, 2, 4, 3, 2, 1, 1)
\end{displaymath}
By the rule proposed above, the Dynkin labels of all the representations we need to compute the index to $O(\tau^2)$ are (1000000), (2000000), (0100000), (0000100), (0000010), (0000020), (0000001), (0000002), (1000010), (1000001) and (0000011). These representations have dimensions 133, 7371, 8645, 1539, 56, 1463, 912, 253935, 6480, 86184, 40755 respectively. This is in agreement with the result shown in \cite{Chacaltana:2017boe}.

For the $E_8$ theory, the largest representation contributing to the SCI of a fixture is 4881384, and this occurs in two fixtures:

\begin{displaymath}
\begin{matrix} \includegraphics[width=76pt]{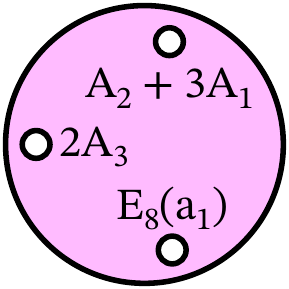}\end{matrix}
\quad\text{and}\quad
\begin{matrix} \includegraphics[width=76pt]{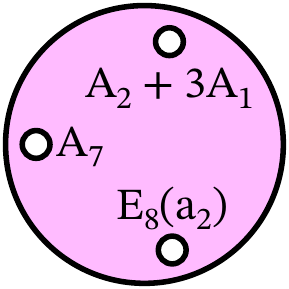}\end{matrix}
\end{displaymath}
From

\begin{displaymath}
P_{A_2+3A_1,2A_3,E_8(a_1)}=(1,3,6,4,3,2,1,3)
\end{displaymath}
we see that the index for the first fixture receives a contribution from the representation $(20000000)=4881384$, and similarly for the second. However, the majority of fixtures require far less computation.

\section{Applications}\label{applications}

\subsection{A Very Special Piece}\label{a_very_special_piece}

Consider the fixture

\begin{displaymath}
 \includegraphics[width=93pt]{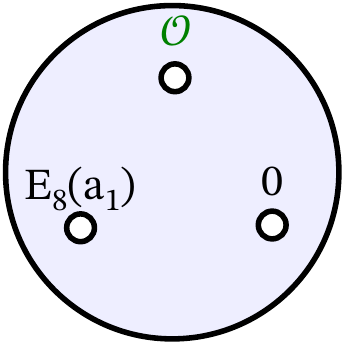}
\end{displaymath}
where $\color{green} \mathcal{O}$ is drawn from the Special Piece of $E_8(a_7)$, i.e., $\color{green}\mathcal{O}$ is one of
\begin{flalign*}
\quad\bullet\;& A_4+A_3         &&\bigl(C(A_4+A_3)=S_5\bigr)&&&\\
\bullet\;& A_5+A_1         &&\bigl(C(A_5+A_1)=S_3\times\mathbb{Z}_2\bigr)&&&\\
\bullet\;& D_5(a_1)+A_2&&\bigl(C(D_5(a_1))=S_4\bigr)&&&\\
\bullet\;& D_6(a_2)        &&\bigl(C(D_6(a_2))=\mathbb{Z}_2\times\mathbb{Z}_2\bigr)&&&\\
\bullet\;& E_6(a_3)+A_1&&\bigl(C(E_6(a_3)+A_1)=S_3\bigr)&&&\\
\bullet\;& E_7(a_5)        &&\bigl(C(E_7(a_5))=\mathbb{Z}_2\bigr)&&&\\
\bullet\;& E_8(a_7).
\end{flalign*}
Luzstig's canonical quotient, $\overline{A}(E_8(a_7))=S_5$. So the Sommers-Achar groups, $C({\color{green}\mathcal{O}})$, associated to the $\color{green}\mathcal{O}$s are various subgroups of $S_5$.

Choosing ${\color{green}\mathcal{O}}=E_8(a_7)$, we get 5 decoupled copies of the ${(E_8)}_{12}$ Minahan-Nemeschansky theory, whose Coulomb branch is 1-dimensional and whose Higgs branch is the 1-instanton moduli space for $E_8$. At the other end, choosing ${\color{green}\mathcal{O}}=A_4+A_3$ yields the $k=5$ Minahan-Nemeschansky,theory, ${(E_8)}_{60}\times {SU(2)}_{124}$, whose Coulomb branch is the $S_5$ quotient of the ${\color{green}\mathcal{O}}=E_8(a_7)$ case, and whose Higgs branch is the $5$-instanton moduli space for $E_8$.

The intermediate cases lie somewhere in between. For instance, ${\color{green}\mathcal{O}}=A_5+A_1$ (Sommers-Achar group $=S_3\times\mathbb{Z}_2$) yields the product of one copy of the $k=3$, ${(E_8)}_{36}\times{SU(2)}_{38}$, Minahan-Nemeschansky theory with one copy of the $k=2$, ${(E_8)}_{24}\times{SU(2)}_{13}$, Minahan-Nemeschansky theory.

If we take one of the above fixtures, and gauge an $E_7$ subgroup of the diagonal $E_8$ (introducing an additional $\tfrac{1}{2}(56)$ to make the $\beta$-function vanish),

\begin{displaymath}
 \includegraphics[width=258pt]{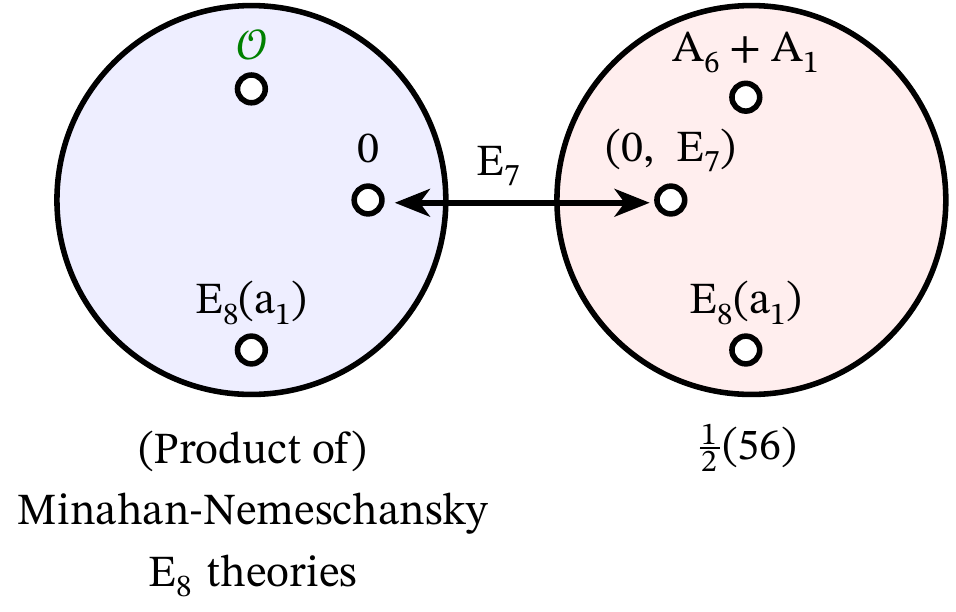}
\end{displaymath}
the global symmetry is the product of the $SU(2)_{k_i}$ which centralize the $E_7\subset \prod_i (E_8)_{k_i}$ stemming from the enhanced $(E_8)_{60}$ of the full puncture (and an additional $SU(2)$ when ${\color{green}\mathcal{O}}$ is non-special). S-dualizing, we obtain

\begin{displaymath}
 \includegraphics[width=243pt]{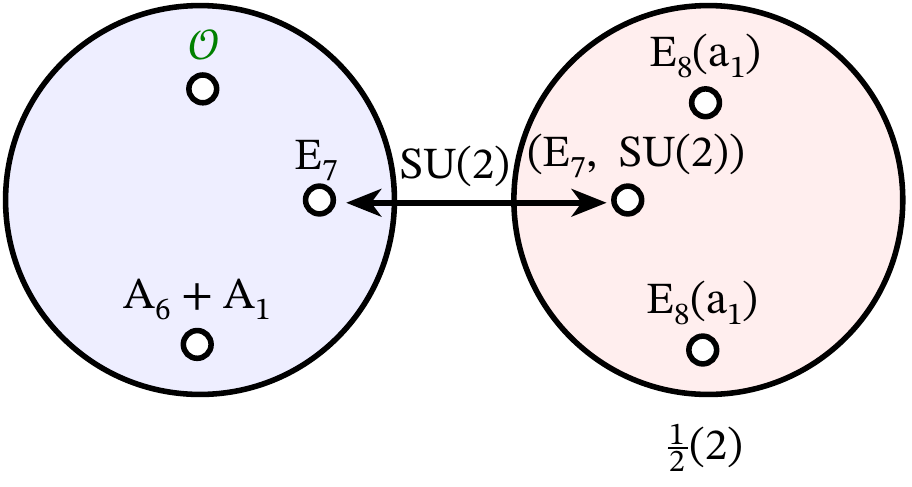}
\end{displaymath}
where the global symmetry of the interacting fixture on the left is the \emph{same} product of $SU(2)$s (times an additional $SU(2)_7$, which we gauge). We can, thus, fill in the levels of the $SU(2)_{k_i}$.

Replacing $E_8(a_1)$ by another distinguished puncture ($E_8(a_2)$, $E_8(a_3)$, $E_8(a_4)$, $E_8(a_5)$, $E_8(a_6)$, $E_8(a_7)$, $E_8(b_4)$ or $E_8(b_6)$), the ${(E_8)}_{60}$ global symmetry associated to the $0$ puncture is not enhanced to a product of $E_8$s, so those fixtures don't follow the same pattern of product SCFTs that this example did.

However, the same pattern arises for the fixture
\begin{displaymath}
 \includegraphics[width=100pt]{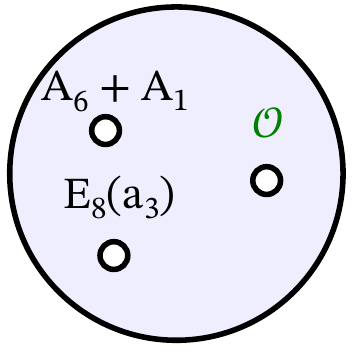}
\end{displaymath}
by letting the puncture ${\color{green}\mathcal{O}}$ vary over the same special piece. When ${\color{green}\mathcal{O}}=E_8(a_7)$, the $SU(2)_{60}$ of the puncture $A_6+A_1$ enhances to $SU(2)^5$. This is the global symmetry enhancement $F_{kn} \to (F_k)^n$ discussed in \S5 of \cite{Chacaltana:2015bna}: as ${\color{green}\mathcal{O}}$ varies over the special piece, the enhanced factor of the global symmetry group is the subgroup of $(F_k)^n$ which is invariant under the action of $C({\color{green}\mathcal{O}})$, which permutes the $n$ copies of $F_k$.

{
\scriptsize
\setlength\LTleft{-.25in}
\renewcommand{\arraystretch}{2.25}

\begin{longtable}{|c|c|c|c|}
\hline
${\color{green}\mathcal{O}}$&$C({\color{green}\mathcal{O}})$&Enhanced Global symmetry&Levels Determined\\
\hline
\endhead
$E_8(a_7)$&1&$\color{red}SU(2)_{60-k_1-k_2-k_3-k_4}\times SU(2)_{k_1}\times SU(2)_{k_2}\times SU(2)_{k_3}\times SU(2)_{k_4}$&$\color{red}{SU(2)_{12}}^5$\\
\hline
$E_7(a_5)$&$\mathbf{Z}_2$&$SU(2)_{13}\times\color{red}SU(2)_{60-k_1-k_2-k_3}\times SU(2)_{k_1}\times SU(2)_{k_2} \times SU(2)_{k_3}$&$SU(2)_{13}\times\color{red}SU(2)_{24}\times {SU(2)_{12}}^3$\\
\hline
$E_6(a_3)+A_1$&$S_3$&$SU(2)_{38}\times\color{red}SU(2)_{60-k_1-k_2}\times SU(2)_{k_1}\times SU(2)_{k_2}$&$SU(2)_{38}\times\color{red}SU(2)_{36}\times {SU(2)_{12}}^2$\\
\hline
$D_6(a_2)$&$\mathbf{Z}_2 \times \mathbf{Z}_2$&${SU(2)_{13}}^2\times\color{red}SU(2)_{60-k_1-k_2}\times SU(2)_{k_1}\times SU(2)_{k_2}$&${SU(2)_{13}}^2\times\color{red}{SU(2)_{24}}^2\times SU(2)_{12}$\\
\hline
$D_5(a_1)+A_2$&$S_4$&$SU(2)_{75}\times\color{red}SU(2)_{60-k}\times SU(2)_{k}$&$SU(2)_{75}\times\color{red}SU(2)_{48}\times SU(2)_{12}$\\
\hline
$A_5+A_1$&$S_3 \times \mathbf{Z}_2$&$SU(2)_{38}\times SU(2)_{13}\times\color{red}SU(2)_{60-k}\times SU(2)_{k}$&$SU(2)_{38}\times SU(2)_{13}\times\color{red}SU(2)_{36}\times SU(2)_{24}$\\
\hline
$A_4+A_3$&$S_5$&$SU(2)_{124}\times SU(2)_{60}$&$SU(2)_{124}\times SU(2)_{60}$\\
\hline
\end{longtable}
}

As in our previous papers \cite{Chacaltana:2015bna,Chacaltana:2017boe}, in this case we can use the action of the Sommers-Achar group on the Higgs branch to fill the missing levels, in the last column of the table.

We have used this action of the Sommers-Achar group on the Higgs branch to fill in the levels for various other families of fixtures which undergo this pattern of global symmetry enhancement.

\subsection{Fun with Minahan-Nemeschansky Theories}\label{fun_with_minahannemeschansky_theories}

There are \emph{no} purely Lagrangian field theories among the Class-S theories of type $E_8$. All of the S-dualities relate gaugings of one non-Lagrangian SCFT to another. The next-best thing would be to replace free hypermultiplets with Minahan-Nemeschansky theories. For instance, we could take two of the the fixtures discussed in \S\ref{a_very_special_piece}, where $\mathcal{O}_1,\mathcal{O}_2$ are drawn from the special piece of $E_8(a_7)$, and gauge the diagonal $E_8$. The S-dual theory

\begin{displaymath}
 \includegraphics[width=212pt]{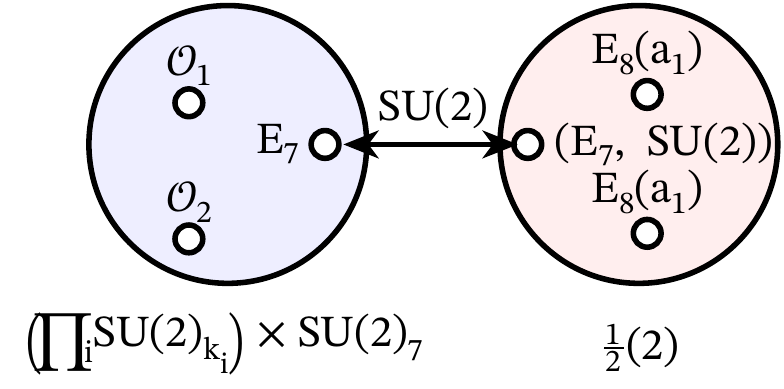}
\end{displaymath}
is an $SU(2)$ gauge theory with a half-hypermultiplet in the fundamental, gauging the ${SU(2)}_7$ subgroup of the global symmetry of the fixture on the left. The remaining global symmetry is the product of the $SU(2)$s that centralized the $E_8$ global symmetry of the Minahan-Nemeschansky theory.

As another example, we can consider a $Spin(13)$ gauge theory, with hypermultiplets in the $2(13)+1(1)$ and the rest of the $\beta$-function condition saturated by coupling to the rank-3 ${(E_8)}_{36}\times {SU(2)}_{38}$ Minahan-Nemeschansky theory.

\begin{displaymath}
 \includegraphics[width=252pt]{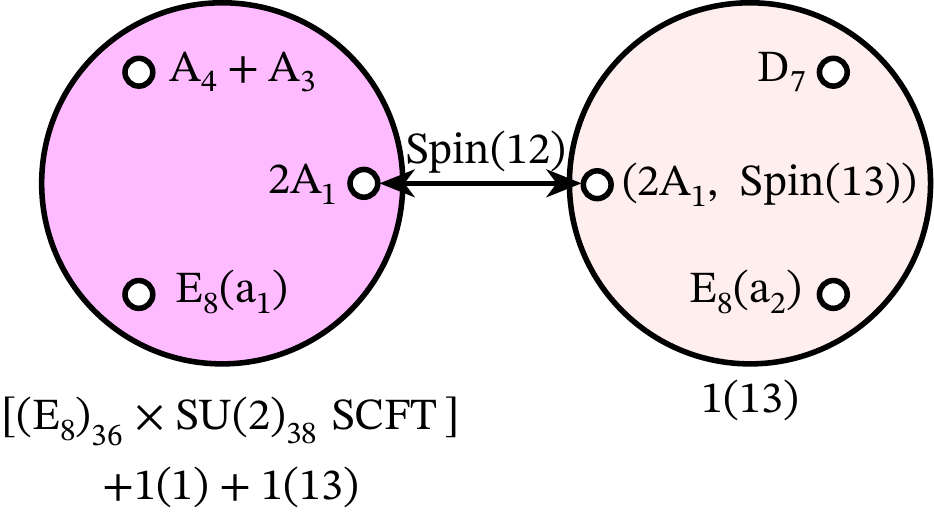}
\end{displaymath}
which is S-dual to

\begin{displaymath}
 \includegraphics[width=252pt]{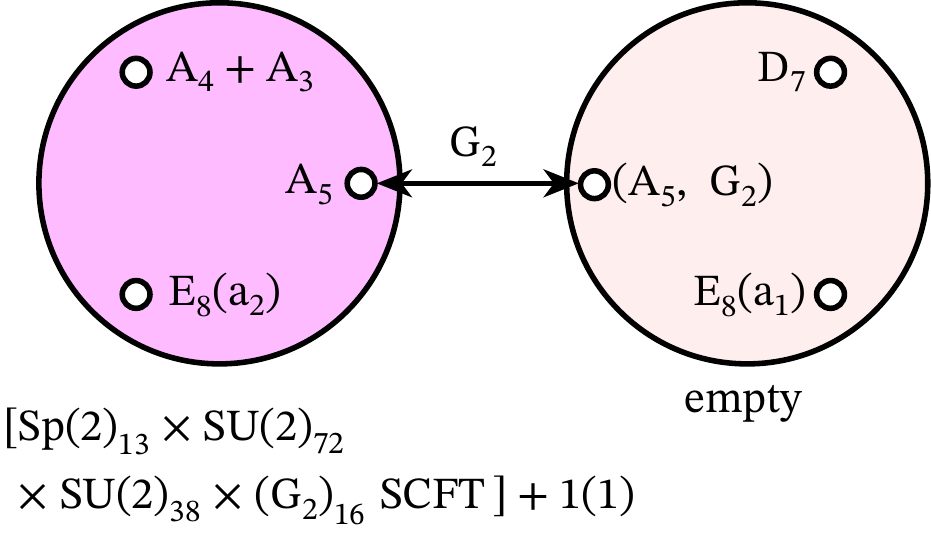}
\end{displaymath}
and

\begin{displaymath}
 \includegraphics[width=252pt]{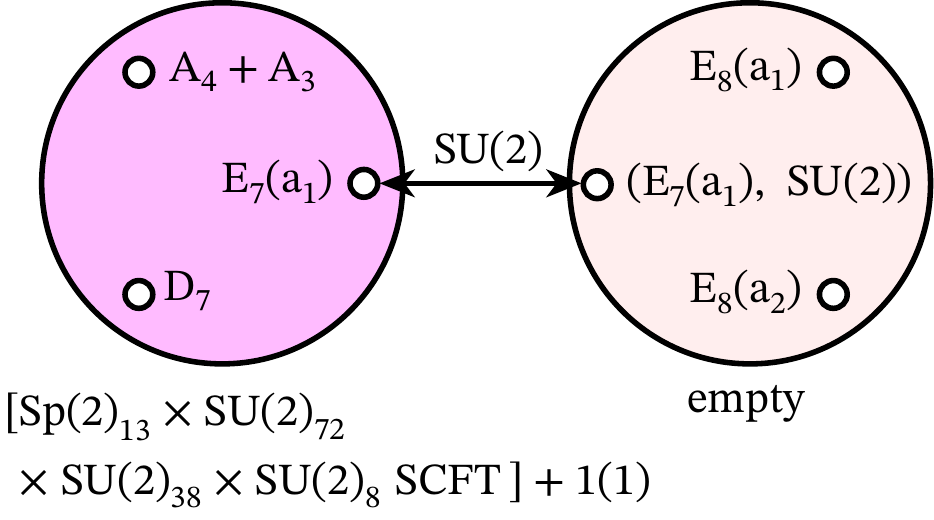}
\end{displaymath}
Using one of the hypermultiplets in the $(13)$ to Higgs $Spin(13)$ to $Spin(12)$, we obtain a $Spin(12)$ gauge theory, with hypermultiplets in the $1(12)+2(1)$, coupling to the same rank-3 ${(E_8)}_{36}\times {SU(2)}_{38}$ Minahan-Nemeschansky theory.

\begin{displaymath}
 \includegraphics[width=252pt]{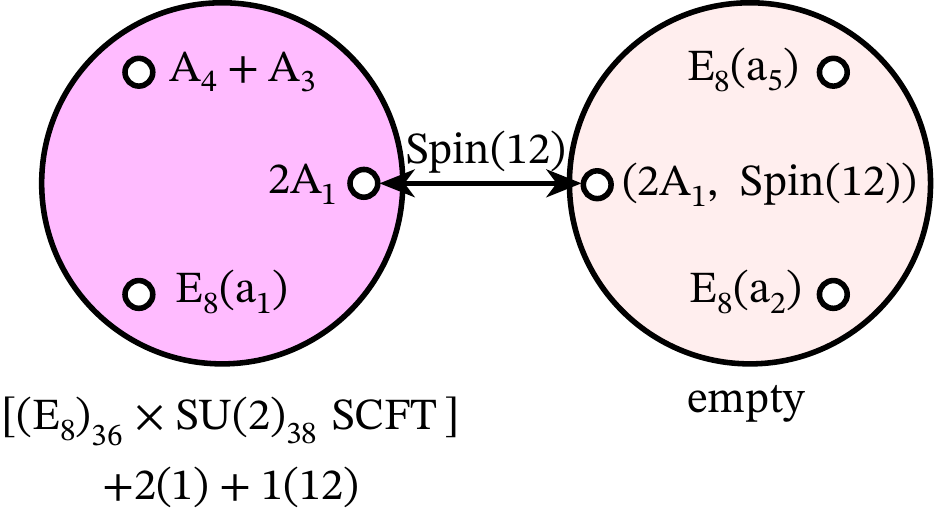}
\end{displaymath}
This is S-dual to

\begin{displaymath}
 \includegraphics[width=252pt]{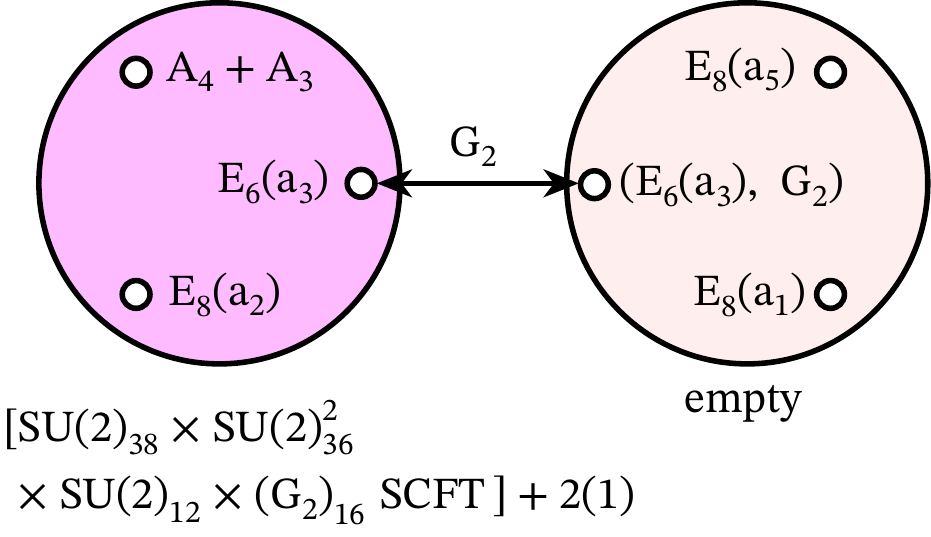}
\end{displaymath}
and

\begin{displaymath}
 \includegraphics[width=252pt]{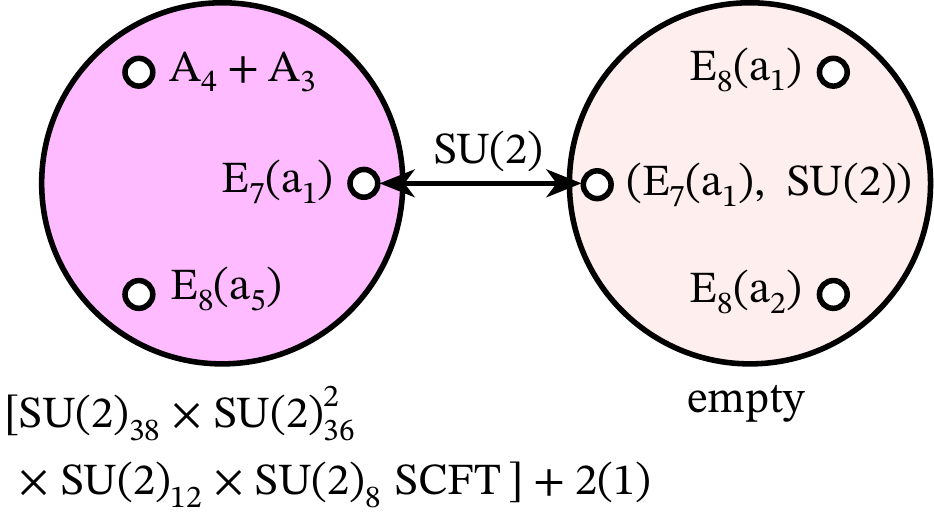}
\end{displaymath}
\subsection{The ${(E_8)}_{20}$ Theory}\label{the__theory_2}

Recently, Giacomelli \cite{Giacomelli:2017ckh} gave a string theory construction of a rank-2 $\mathcal{N}=2$ SCFT with global symmetry ${(E_8)}_{20}$, $n_4=n_{10}=1$ and $(n_h,n_v)=(72,26)$. This SCFT is realized in the $A_9$ theory \cite{Chacaltana:2010ks} as the fixture

\begin{displaymath}
 \includegraphics[width=93pt]{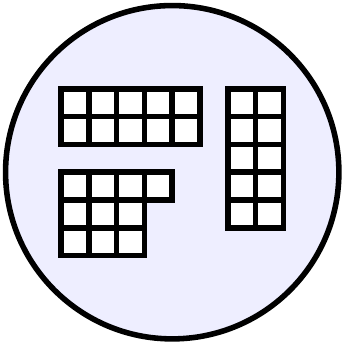}
\end{displaymath}
and in the $D_6$ theory \cite{Chacaltana:2011ze} as the fixture

\begin{displaymath}
 \includegraphics[width=138pt]{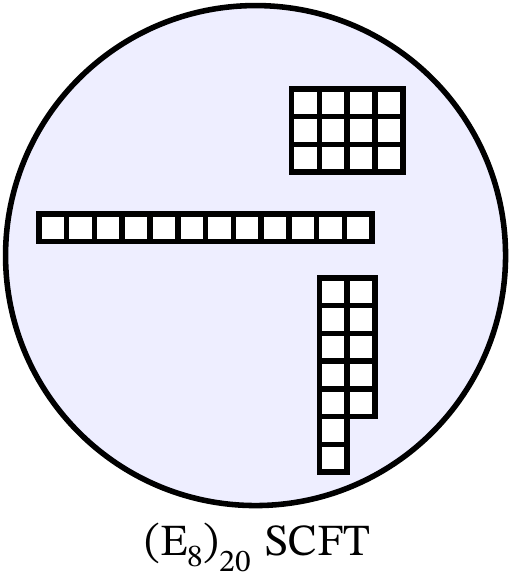}
\end{displaymath}
The same SCFT appeared at a factor in product theories in the $E_7$ theory \cite{Chacaltana:2017boe}.

\begin{displaymath}
\begin{matrix} \includegraphics[width=93pt]{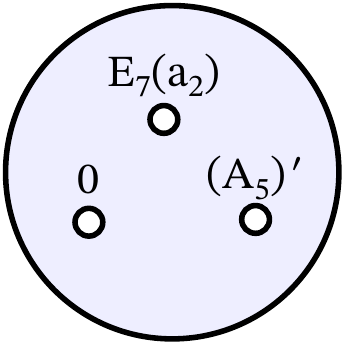}
&\qquad& \includegraphics[width=93pt]{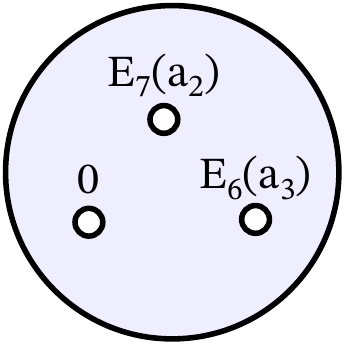}\\
\begin{gathered}
[{(E_8)}_{20}\, \text{SCFT}]\\
\times\\
[{(E_7)}_{16}\times {SU(2)}_9\,\text{SCFT}]
\end{gathered}&\qquad&
\begin{gathered}
[{(E_8)}_{20}\, \text{SCFT}]\\
\times\\
{[{(E_7)}_{8}\,\text{SCFT}]}^2
\end{gathered}
\end{matrix}
\end{displaymath}
It also appears here in the $E_8$ theory, again as a product, but this time as three copies of itself.

\begin{displaymath}
\begin{matrix} \includegraphics[width=93pt]{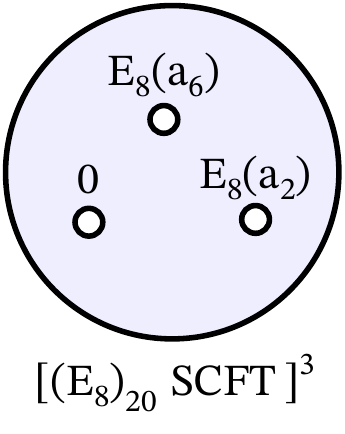}\end{matrix}
\end{displaymath}
While the Coulomb branch is 2-dimensional, the Higgs branch of the ${(E_8)}_{20}$ theory is also interesting: it has $\text{dim}_\mathbb{H}=46$ and is conjectured\footnote{Note that $c=(2n_v+n_h)/12$ saturates the unitarity bound\eqref{unitarityklowerbound}
which implies the absence of flavour-singlet $\hat{B}_2$ multiplets in the spectrum \cite{Beem:2013sza}.}  to be the (closure of the) $2A_1$ nilpotent orbit of $E_8$.

\subsection{Theories with $E_8$ Global Symmetry}\label{theories_with__global_symmetry}

As already mentioned, a fixture with a pair of distinguished punctures and one full puncture has a manifest ${(E_8)}_{60}$ global symmetry. There are 32 fixtures where this is, in fact, the global symmetry. There are 5 more where the symmetry is enhanced.

We have already discussed the case of $(E_8(a_7),E_8(a_1))$, where the enhanced symmetry is ${(E_8)}_{12}^5$ and the fixture is actually 5 copies of the ${(E_8)}_{12}$ Minahan-Nemeschansky theory, and the case of $(E_8(a_6),E_8(a_2))$, where we get three copies of the ${(E_8)}_{20}$ SCFT.

The remaining three cases are also interesting.

\begin{displaymath}
 \includegraphics[width=165pt]{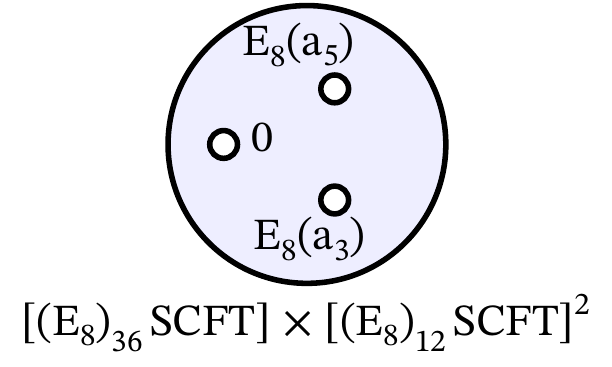}
\end{displaymath}
is the rank-5 ${(E_8)}_{36}$ theory (first found in \S4.3.3 of \cite{Chacaltana:2017boe}) plus two copies of the ${(E_8)}_{12}$ theory. The fixture

\begin{equation}
 \includegraphics[width=83pt]{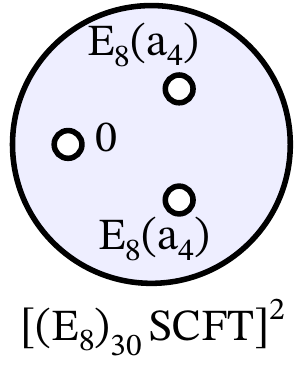}
\label{E830squared}\end{equation}
has an enhanced ${(E_8)}_{60-k}\times {(E_8)}_{k}$ global symmetry and $(n_h,n_v)=(256,120)$. This gives $c=124/3$. If the theory were indecomposable (possessing a single stress tensor), then this would violate the unitarity bound \cite{Beem:2013sza}

\begin{equation}
k_i \geq \frac{24 \kappa_{F_i}c}{\dim(F_i)+12 c}
\label{unitarityklowerbound}\end{equation}
for one or the other of the two $E_8$s. The resolution is that \eqref{E830squared} must be a product of two SCFTs, with global symmetries $F_i = {(E_8)}_{k_i}$ and Weyl anomaly coefficients $c_i$ satisfying $k_1+k_2=60$ and $c_1+c_2=124/3$. The fact that \eqref{E830squared} has graded Coulomb branch dimensions $n_3=n_5=n_9=n_{15}=2$ excludes obvious candidates, like the ${(E_8)}_{12}$, the ${(E_8)}_{20}$ and the ${(E_8)}_{36}$ theories, from being one of the factors in the product. Instead, it strongly suggests\footnote{A very similar phenomenon occurs with the fixture

\begin{displaymath}
 \includegraphics[width=137pt]{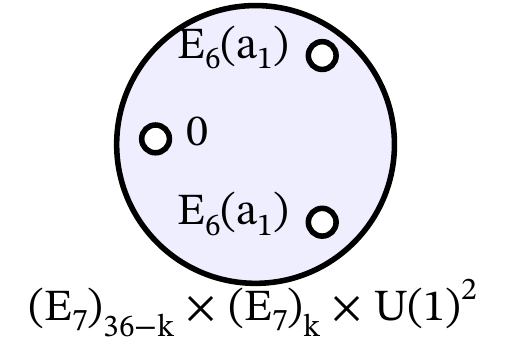}
\end{displaymath}
in the $E_7$ theory. For the same reason, this must be a product SCFT and, in fact, is two copies of a rank-3 theory with global symmetry ${(E_7)}_{18}\times U(1)$, $(n_h,n_v)=(70, 31)$ and $n_3=n_5=n_9=1$.}  that \eqref{E830squared} is the product of two copies of a rank-4 SCFT, with global symmetry ${(E_8)}_{30}$, $(n_h,n_v)=(128,60)$ and graded Coulomb branch dimensions $n_3=n_5=n_9=n_{15}=1$. This choice saturates the bound \eqref{unitarityklowerbound} and $(3,5,9,15)$ is an allowed 4-tuple of graded Coulomb branch dimensions \cite{Caorsi:2018zsq}.

This ${(E_8)}_{30}$ SCFT is realized in the $A_{14}$ theory as the fixture

\begin{displaymath}
 \includegraphics[width=108pt]{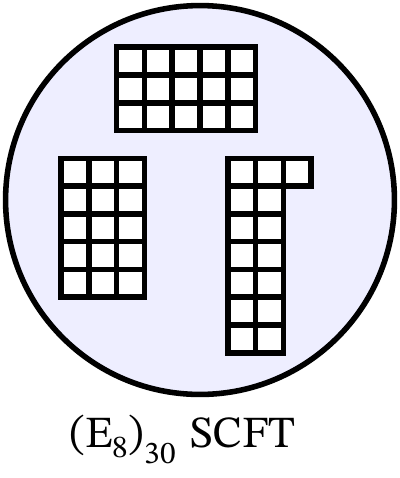}
\end{displaymath}
where only an ${SU(5)}_{30}\times {SU(3)}_{30}\times {SU(2)}_{30}\times U(1)$ subgroup of the global symmetry is manifest.

Finally, we turn to

\begin{displaymath}
 \includegraphics[width=165pt]{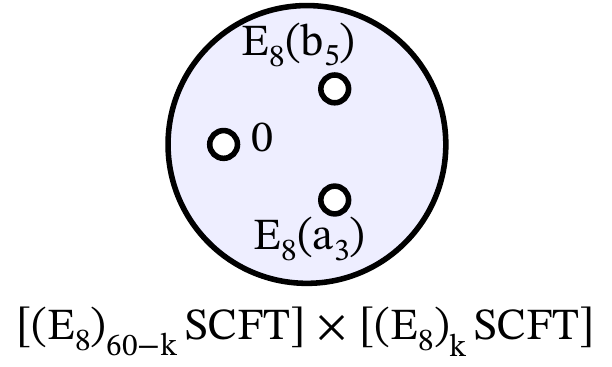}
\end{displaymath}
Again, the unitarity bound \eqref{unitarityklowerbound} would be violated, unless this is a product theory. But here, we are not so lucky in identifying candidate factors for the product. One possibility is that one of the factors is the ${(E_8)}_{12}$ SCFT, and the other is a rank-11 SCFT with global symmetry ${(E_8)}_{48}$, $(n_h,n_v)=(344, 235)$ and $n_3=n_6=n_8=n_{14}=n_{24}=1$, $n_4=n_{12}=n_{18}=2$. But there are several others, with no good way to decide between them.

\subsection{More Product Theories from Enhanced ${(E_8)}_{60}$}\label{more_product_theories_from_enhanced_}

There are 8 more theories, where the ${(E_8)}_{60}$ symmetry of the full puncture is enhanced to ${(E_8)}_{60-k}\times {(E_8)}_{k}$. Again, in all these cases, the unitarity bound \eqref{unitarityklowerbound} requires that these are actually product SCFTs.

{
\footnotesize
\renewcommand{\arraystretch}{2.25}

\begin{longtable}{|c|c|c|c|c|}
\hline
\#&Fixture&Graded Coulomb Branch Dims&$(n_h,n_v)$&Global Symmetry\\
\hline
\endhead
1&$\begin{matrix} \includegraphics[width=76pt]{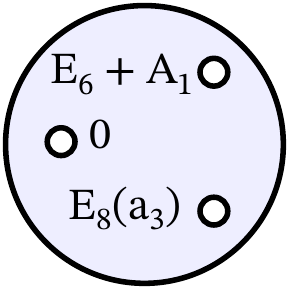}\end{matrix}$&$\{0,1,0,0,2,2,0,0,3,1,0,2,0,1,0\}$&$(410, 270)$&${(E_8)}_{60-k}\times{(E_8)}_{k}\times{SU(2)}_{26}$\\
\hline
2&$\begin{matrix} \includegraphics[width=76pt]{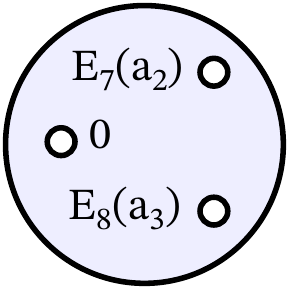}\end{matrix}$&$\{0,1,1,0,2,2,0,0,2,1,0,2,0,1,0\}$&$(393, 254)$&${(E_8)}_{60-k}\times{(E_8)}_{k}\times{SU(2)}_{9}$\\
\hline
3&$\begin{matrix} \includegraphics[width=76pt]{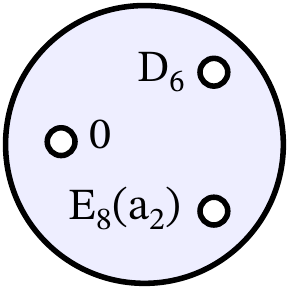}\end{matrix}$&$\{0,0,2,0,0,1,0,3,0,1,0,0,1,0,0\}$&$(294, 152)$&${(E_8)}_{60-k}\times{(E_8)}_{k}\times{Sp(2)}_{11}$\\
\hline
4&$\begin{matrix} \includegraphics[width=76pt]{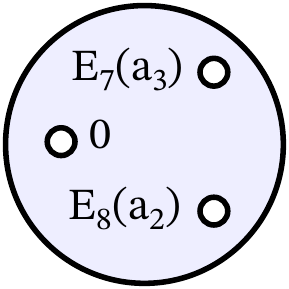}\end{matrix}$&$\{0,0,2,1,0,1,0,2,0,1,0,0,1,0,0\}$&$(282, 142)$&${(E_8)}_{60-k}\times{(E_8)}_{k}\times{SU(2)}_{10}$\\
\hline
5&$\begin{matrix} \includegraphics[width=76pt]{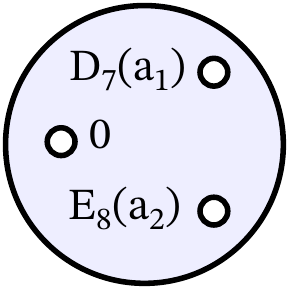}\end{matrix}$&$\{0,0,2,0,0,1,0,2,0,1,0,0,1,0,0\}$&$(272, 133)$&${(E_8)}_{60-k}\times{(E_8)}_{k}\times U(1)$\\
\hline
6&$\begin{matrix} \includegraphics[width=76pt]{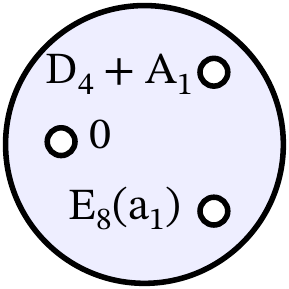}\end{matrix}$&$\{0,0,0,0,2,1,0,0,1,1,0,2,0,1,0\}$&$(361, 204)$&${(E_8)}_{60-k}\times{(E_8)}_{k}\times{Sp(3)}_{19}$\\
\hline
7&$\begin{matrix} \includegraphics[width=76pt]{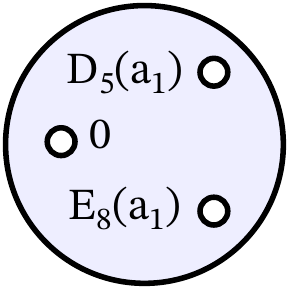}\end{matrix}$&$\{0,0,0,0,2,1,1,0,1,1,0,1,0,1,0\}$&$(340, 186)$&${(E_8)}_{60-k}\times{(E_8)}_{k}\times{SU(4)}_{18}$\\
\hline
8&$\begin{matrix} \includegraphics[width=76pt]{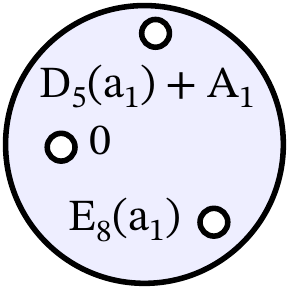}\end{matrix}$&$\{0,0,0,0,2,1,0,0,1,1,0,1,0,1,0\}$&$(320, 169)$&${\begin{gathered}{(E_8)}_{60-k}\times{(E_8)}_{k}\\ \times{SU(2)}_{16}\times{SU(2)}_{112}\end{gathered}}$\\
\hline
\end{longtable}
}
\noindent
with Graded Coulomb Branch dimensions $\{n_2,n_3,n_4,n_5,n_6,n_8,n_9,n_{10},n_{12},n_{14},n_{15},n_{18},n_{20},n_{24},n_{30}\}$.

Some of these have obvious candidates for the factors in the product. \#7 is the product of the ${(E_8)}_{12}$ Minahan-Nemeschansky SCFT with a rank-7 ${(E_8)}_{48}\times {SU(4)}_{18}$ SCFT with $(n_h,n_v)=(300, 175)$ which is realized in the $A_{23}$ theory as the fixture

\begin{displaymath}
 \includegraphics[width=138pt]{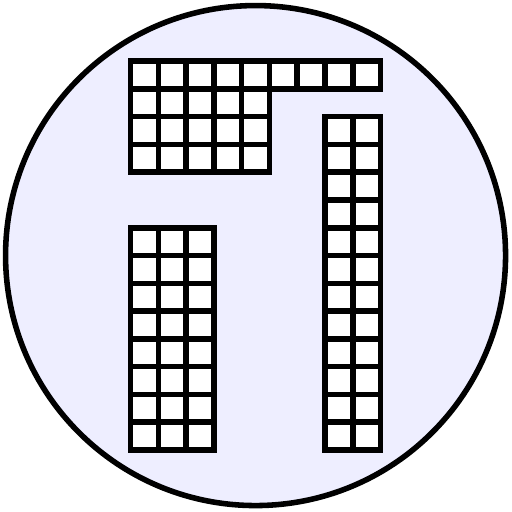}
\end{displaymath}
\#4 is the product of the ${(E_8)}_{20}$ SCFT discussed above with a rank-5 ${(E_8)}_{40}\times {SU(2)}_{10}$ SCFT, with $(n_h,n_v)=(210, 116)$, which is realized in the $A_{19}$ theory as the fixture

\begin{displaymath}
 \includegraphics[width=123pt]{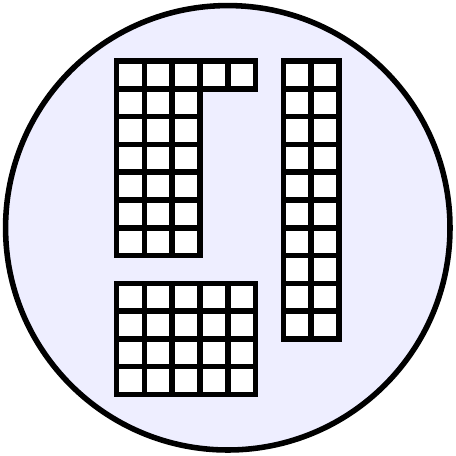}
\end{displaymath}
It is likely that \#s 3,5 are products of the ${(E_8)}_{20}$ theory with another SCFT and that the rest are products of the ${(E_8)}_{12}$ SCFT with another theory. But we have not found suitable candidates for alternate realizations of the other factor in the product for these cases.

A similar phenomenon occurs \cite{Ergun:2018xxx} in the $E_7$ theory: fixtures where the ${(E_7)}_{36}$ symmetry of the full puncture is enhanced to ${(E_7)}_{36-k}\times {(E_7)}_{k}$ and hence (by the unitarity bound) must be product SCFTs.

This seems to be a fruitful problem for further investigation: given that we know an SCFT is a product, how to determine the factors in the product. One piece of very useful information (which usually suffices when the rank is low) is knowing what are the allowed $n$-tuples of Coulomb branch scaling dimensions \cite{Caorsi:2018zsq,Argyres:2018urp}. That certainly narrows the possibilities but is not --- at least in the present cases --- quite sufficient to nail things down. 

\subsection{Compactifications of 6D Conformal Matter Theories}\label{compactifications_of_6d_conformal_matter_theories}

Another source of 4D $\mathcal{N}=2$ SCFTs is the compactification of 6D $(1, 0)$ SCFTs on a torus \cite{Ohmori:2015pua,Ohmori:2015pia,DelZotto:2015rca}. The compactification of the $(E_8,G_2)$ conformal matter theory \cite{DelZotto:2014hpa} yields a 4D $\mathcal{N}=2$ SCFT with $(E_8)_{36}\times (G_2)_{16}$ global symmetry which was realized \cite{DelZotto:2015rca,Chacaltana:2017boe} as the $E_7$ fixture

\begin{displaymath}
\begin{matrix} \includegraphics[width=113pt]{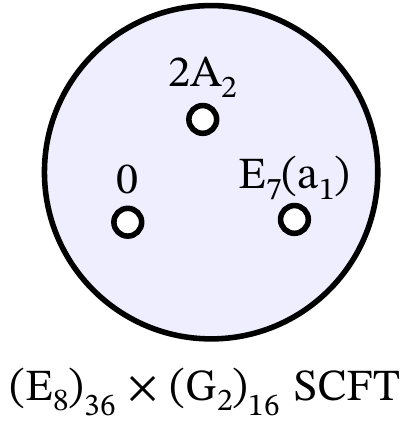}\end{matrix}
\end{displaymath}
It also appeared in \cite{Giacomelli:2017ckh} as $D_2^{20}(E_8)$. Here, we see it arising as a factor in a product theory, for two fixtures\footnote{The two fixtures are related by the usual Sommers-Achar action on the $E_8$ Minahan-Nemeschansky theories. $A_5$ and $E_6(a_3)$ lie in the same special piece, with $d(A_5)=(D_6(a_1),\mathbb{Z}_2)$ and $d(E_6(a_3))=D_6(a_1)$; so the fixture with the $A_5$ puncture contains the rank-2 Minahan-Nemeschansky theory, where the fixture with the $E_6(a_3)$ puncture contains two copies of the rank-1 Minahan-Nemeschansky theory.} :

\begin{displaymath}
\begin{matrix} \includegraphics[width=93pt]{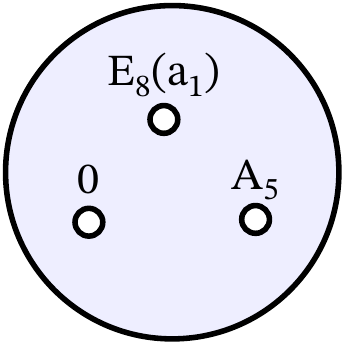}
&\qquad& \includegraphics[width=93pt]{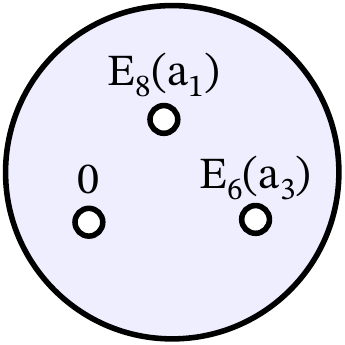}\\
\begin{gathered}
[{(E_8)}_{36}\times {(G_2)}_{16}\, \text{SCFT}]\\
\times\\
[{(E_8)}_{24}\times {SU(2)}_{13}\,\text{SCFT}]
\end{gathered}&\qquad&
\begin{gathered}
[{(E_8)}_{36}\times {(G_2)}_{16}\, \text{SCFT}]\\
\times\\
{[{(E_8)}_{12}\,\text{SCFT}]}^2
\end{gathered}
\end{matrix}
\end{displaymath}
Similarly, the compactification of the $(E_7,SO(7))$ conformal matter theory on a torus yields a 4D $\mathcal{N}=2$ SCFT with $(E_7)_{24}\times Spin(7)_{16}$ global symmetry. It is realized in the $E_8$ theory as the mixed fixture (accompanied by $15$ hypermultiplets)

\begin{displaymath}
\begin{matrix} \includegraphics[width=131pt]{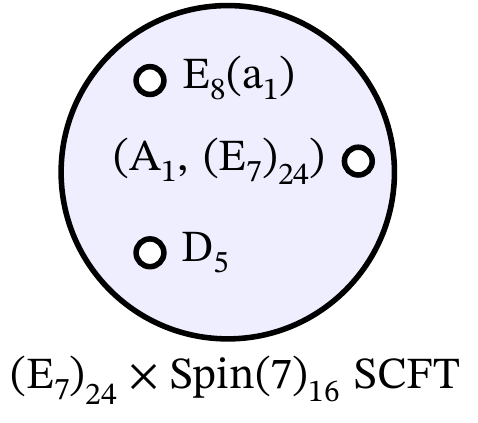}\end{matrix}
\end{displaymath}
and also as the irregular fixture

\begin{displaymath}
\begin{matrix} \includegraphics[width=129pt]{E7Spin7new1}\end{matrix}
\end{displaymath}

\section*{Acknowledgements}
We would like to thank Mario Martone, Philip Argyres and Sergio Cecotti for helpful discussions. The work of J.D. and Y.Z.~was supported in part by the National Science Foundation under Grant No. PHY-1620610. J.D.~would also like to thank the organizers of the Pollica Summer Workshop 2017, partly supported by the ERC STG grant 306260, and the Simons Center for Geometry and Physics for their gracious hospitality during an early stage of this work.

%J.D. and A.T. would like to thank the Aspen Center for Physics, which is supported by National Science Foundation grant PHY-1066293,  for hospitality during the workshop ``Superconformal Field Theories in $d \geq 4$", where the final stages of this work was completed. O.C. would like to thank the Johns Hopkins University, and especially Jared Kaplan, for hospitality while this work was completed. O.C. would also like to thank the Aspen Center for Physics, supported by NSF grant PHY-1066293, and acknowledge kind support by a grant from the Simons Foundation, for hospitality during the 2016 Summer Program of the ACP, while this work was completed. O.C. also gratefully acknowledges support from the Simons Center for Geometry and Physics, Stony Brook University, for hospitality at the 2016 Simons Summer Workshop in Mathematics and Physics, where part of this work was completed.
\begin{appendices}

\section{Appendix: Embeddings of $SU(2)$ in $E_8$}\label{appendix_embeddings_of__in_}
{
\footnotesize
\renewcommand{\arraystretch}{2.25}

\begin{longtable}{|c|c|l|}
\hline
Bala-Carter&$\mathfrak{f}$&$248$\\
\hline 
$A_1$&$\mathfrak{e}_7$&$(3;1)+(2;56)+(1;133)$\\
\hline
$2A_1$&$\mathfrak{so}(13)$&$(1;78)+(2;64)+(3;1)+(3;13)$\\
\hline
$3A_1$&$\mathfrak{f}_4\times \mathfrak{su}(2)$&$(1;1,3)+(1;52,1)+(2;26,2)+(3;1,1)+(3;26,1)+(4;1,2)$\\
\hline
$A_2$&$\mathfrak{e}_6$&$(1;78)+(3;1)+(3;27)+(3;\overline{27})+(5;1)$\\
\hline
$4A_1$&$\mathfrak{sp}(4)$&$(1;36)+(2;48)+(3;1)+(3;27)+(4;8)$\\
\hline
$A_2+A_1$&$\mathfrak{su}(6)$&${\begin{aligned}&(1;35)+(2;6)+(2;\overline{6})+(2;20)+2(3;1)+(3;15)+(3;\overline{15})+(4;6)\\&+(4;\overline{6})+(5;1)\end{aligned}}$\\
\hline
$A_2+2A_1$&$\mathfrak{so}(7)\times \mathfrak{su}(2)$&${\begin{aligned}&(1;1,3)+(1;21,1)+(2;8,4)+(3;7,3)+(3;1,5)+(3;1,1)+(4;8,2)\\&+(5;1,3)\end{aligned}}$\\
\hline
$A_3$&$\mathfrak{so}(11)$&$(1;55)+(3;1)+(4;32)+(5;11)+(7;1)$\\
\hline
$A_2+3A_1$&$\mathfrak{g}_2\times \mathfrak{su}(2)$&$(1;14,1)+(1;1,3)+(2;14,2)+(3;27,1)+(3;1,1)+(4;7,2)+(5;7,1)$\\
\hline
$2A_2$&$(\mathfrak{g}_2)^2$&$(1;14,1)+(1;1,14)+(3;7,7)+(3;1,1)+(5;1,7)+(5;7,1)$\\
\hline
$2A_2+A_1$&$\mathfrak{g}_2\times \mathfrak{su}(2)$&${\begin{aligned}&(1;1,3)+(1;14,1)+(2;1,4)+(2;7,2)+2(3;1,1)+(3;7,3)\\&+(4;1,2)+(4;7,2)+(5;1,3)+(5;7,1)+(6;1,2)\end{aligned}}$\\
\hline
$A_3+A_1$&$\mathfrak{so}(7)\times \mathfrak{su}(2)$&${\begin{aligned}&(1;1,3)+(1;21,1)+(2;7,2)+2(3;1,1)+(3;8,1)+(4;8,2)\\&+(4;1,2)+(5;8,1)+(5;7,1)+(6;1,2)+(7;1,1)\end{aligned}}$\\
\hline
$D_4(a_1)$&$\mathfrak{so}(8)$&${\begin{aligned}&(1;28)+3(3;1)+(3;8_v)+(3;8_c)+(3;8_s)+(5;1)\\&+(5;8_v)+(5;8_c)+(5;8_s)+2(7;1)\end{aligned}}$\\
\hline
$D_4$&$\mathfrak{f}_4$&$(1;52)+(3;1)+(7;26)+(11;1)$\\
\hline
$2A_2+2A_1$&$\mathfrak{sp}(2)$&$(1;10)+(2;20)+(3;1)+(3;5)+(3;14)+(4;16)+(5;10)+(6;4)$\\
\hline
$A_3+2A_1$&$\mathfrak{sp}(2)\times \mathfrak{su}(2)$&
${\begin{aligned}&(1;10,1)+(1;1,3)+(2;4,3)+(2;1,2)+2(3;1,1)\\&+(3;5,1)+(3;4,2)+(4;1,2)+(4;4,1)+(4;5,2)\\ &+(5;4,2)+(5;1,3)+(6;4,1)+(6;1,2)+(7;1,1)\end{aligned}}$
\\
\hline
$D_4(a_1)+A_1$&$\mathfrak{su}(2)^3$&${\begin{aligned}&(1;1,1,3)+(1;1,3,1)+(1;3,1,1)+(2;2,2,2)+(2;2,1,1)\\&+(2;1,1,2)+(2;1,2,1)+4(3;1,1,1)+(3;1,2,2)+(3;2,2,1)\\&+(3;2,1,2)+2(4;2,1,1)+2(4;1,2,1)+2(4;1,1,2)+(5;1,1,1)\\&+(5;2,2,1)+(5;1,2,2)+(5;2,1,2)+(6;2,1,1)+(6;1,2,1)\\&+(6;1,1,2)+2(7;1,1,1)\end{aligned}}$\\
\hline
$A_3+A_2$&$\mathfrak{sp}(2)\times \mathfrak{u}(1)$&${\begin{aligned}&(1;10)_0+(1;1)_0+(2;4)_{-1/2}+(2;4)_{1/2}+(3;1)_{-2}+(3;1)_2\\&+2(3;1)_0+(3;5)_1+(3;5)_{-1}+(3;1)_1+(3;1)_{-1}+(4;4)_{3/2}\\&+(4;4)_{1/2}+(4;4)_{-1/2}+(4;4)_{-3/2}+(5;1)_0+(5;5)_0+(5;1)_1\\&+(5;1)_{-1}+(6;4)_{-1/2}+(6;4)_{1/2}+(7;1)_0+(7;1)_1+(7;1)_{-1}\end{aligned}}$\\
\hline
$A_4$&$\mathfrak{su}(5)$&${\begin{aligned}&(1;24)+(3;1)+(3;5)+(3;\overline{5})+(5;1)+(5;10)\\&+(5;\overline{10})+(7;1)+(7;5)+(7;\overline{5})+(9;1)\end{aligned}}$\\
\hline
$A_3+A_2+A_1$&$\mathfrak{su}(2)^2$&${\begin{aligned}&(7;5,1)+(5;7,1)+(5;3,1)+(6;3,2)+(4;7,2)+(3;9,1)\\&+(3;5,1)+(3;1,1)+(2;5,2)+(1;3,1)+(1;1,3)\end{aligned}}$\\
\hline
$D_4+A_1$&$\mathfrak{sp}(3)$&$(1;21)+(2;14')+2(3;1)+(6;6)+(7;14)+(8;6)+(11;1)$\\
\hline
$D_4(a_1)+A_2$&$\mathfrak{su}(3)$&$(1;8)+(3;1)+(3;27)+(5;10)+(5;\overline{10})+(7;8)$\\
\hline
$A_4+A_1$&$\mathfrak{su}(3)\times \mathfrak{u}(1)$&${\begin{aligned}&(1;1)_0+(1;8)_0+(2;3)_{-5}+(2;\overline{3})_5+(2;1)_3+(2;1)_{-3}+2(3;1)_0\\&+(3;3)_{-2}+(3;\overline{3})_2+(4;1)_3+(4;1)_{-3}+(4;3)_1+(4;\overline{3})_{-1}+(5;1)_0\\&+(5;1)_6+(5;1)_{-6}+(5;3)_4+(5;\overline{3})_{-4}+(6;3)_1+(6;\overline{3})_{-1}+(6;1)_3\\&+(6;1)_{-3}+(7;1)_0+(7;3)_{-2}+(7;\overline{3})_2+(8;1)_3+(8;1)_{-3}+(9;1)_0\end{aligned}}$\\
\hline
$2A_3$&$\mathfrak{sp}(2)$&${\begin{aligned}&(8;4)+(7;5)+(7;1)+(6;4)+(5;10)+(4;16)+(3;5)+(3;1)+(2;4)\\&+(1;10)\end{aligned}}$\\
\hline
$D_5(a_1)$&$\mathfrak{su}(4)$&${\begin{aligned}&(1;15)+(2;4)+(2;\overline{4})+2(3;1)+(3;6)+(5;1)+(6;4)+(6;\overline{4})\\&+2(7;1)+(7;6)+(8;4)+(8;\overline{4})+(9;1)+(11;1)\end{aligned}}$\\
\hline
$A_4+2A_1$&$\mathfrak{su}(2)\times \mathfrak{u}(1)$&${\begin{aligned}&(9;1)_0+(8;2)_1+(8;2)_{-1}+(7;1)_0+(7;1)_4+(7;1)_{-4}+(7;1)_2+(7;1)_{-2}\\&+(6;2)_1+(6;2)_{-1}+(6;2)_3+(6;2)_{-3}+(5;1)_0+(5;1)_2+(5;1)_{-2}\\&+(5;3)_2+(5;3)_{-2}+(4;2)_1+(4;2)_{-1}+(4;2)_3+(4;2)_{-3}+(3;1)_2\\&+(3;1)_{-2}+2(3;1)_0+(3;1)_4+(3;1)_{-4}+(3;3)_0+(2;2)_5+(2;2)_{-5}\\&+(2;2)_1+(2;2)_{-1}+(1;1)_0+(1;3)_0\end{aligned}}$\\
\hline
$A_4+A_2$&$\mathfrak{su}(2)^2$&${\begin{aligned}&(1;1,3)+(1;3,1)+(3;1,1)+(3;2,6)+(3;1,5)+(5;1,7)+(5;1,3)\\&(5;2,2)+(7;1,5)+(7;2,4)+(9;1,3)\end{aligned}}$\\
\hline
$A_5$&$\mathfrak{g}_2\times \mathfrak{su}(2)$&${\begin{aligned}&(1;1,3)+(1;14,1)+(3;1,1)+(4;1,2)+(5;7,1)\\&+(6;7,2)+(7;1,1)+(9;7,1)+(10;1,2)+(11;1,1)\end{aligned}}$\\
\hline
$D_5(a_1)+A_1$&$\mathfrak{su}(2)^2$&${\begin{aligned}&(1;1,3)+(1;3,1)+(2;2,5)+(3;1,5)+2(3;1,1)+(4;2,1)+(5;1,3)\\&+(6;2,3)+(7;1,5)+(7;1,3)+(8;2,3)+(9;1,3)+(11;1,1)\end{aligned}}$\\
\hline
$A_4+A_2+A_1$&$\mathfrak{su}(2)$&${\begin{aligned}&(9;3)+(8;4)+(7;5)+(6;2)+(6;4)+(5;3)+(5;7)\\&+(4;2)+(4;6)+2(3;1)+(3;5)+(2;6)+(1;3)\end{aligned}}$\\
\hline
$D_4+A_2$&$\mathfrak{su}(3)$&${\begin{aligned}&(1;8)+2(3;1)+(3;6)+(3;\overline{6})+(5;1)+(5;3)+(5;\overline{3})+(7;3)\\&+(7;\overline{3})+(7;8)+(9;3)+(9;\overline{3})+(11;1)\end{aligned}}$\\
\hline
$E_6(a_3)$&$\mathfrak{g}_2$&$(1;14)+3(3;1)+(5;1)+2(5;7)+(7;1)+(7;7)+(9;1)+(9;7)+2(11;1)$\\
\hline
$D_5$&$\mathfrak{so}(7)$&$(1;21)+(3;1)+(5;8)+(7;1)+(9;7)+(11;1)+(11;8)+(15;1)$\\
\hline
$A_4+A_3$&$\mathfrak{su}(2)$&${\begin{aligned}&(10;2)+(9;3)+(8;4)+(7;1)+(7;5)+(6;2)+(6;4)\\&+2(5;3)+(4;2)+(4;6)+(3;5)+(3;1)+(2;4)+(1;3)\end{aligned}}$\\
\hline
$A_5+A_1$&$\mathfrak{su}(2)^2$&${\begin{aligned}&(1;1,3)+(1;3,1)+(2;4,1)+2(3;1,1)+(4;1,2)+(4;2,1)\\&+(5;3,1)+(5;2,2)+(6;2,1)+(6;3,2)+(7;1,1)+(7;2,2)\\&+(8;2,1)+(9;3,1)+(10;1,2)+(10;2,1)+(11;1,1)\end{aligned}}$\\
\hline
$D_5(a_1)+A_2$&$\mathfrak{su}(2)$&${\begin{aligned}&(11;1)+(10;2)+(9;3)+(8;4)+(8;2)+(7;3)+(7;1)\\&+(6;4)+(6;2)+2(5;3)+2(4;2)+2(3;1)+(3;5)+(2;4)+(1;3)\end{aligned}}$\\
\hline
$D_6(a_2)$&$\mathfrak{su}(2)^2$&${\begin{aligned}&(1;1,3)+(1;3,1)+3(3;1,1)+2(4;2,1)+2(4;1,2)\\&+(5;1,1)+(5;2,2)+(6;1,2)+(6;2,1)+3(7;1,1)+(7;2,2)\\&+(8;1,2)+(8;2,1)+(9;1,1)+(10;1,2)+(10;2,1)+2(11;1,1)\end{aligned}}$\\
\hline
$E_6(a_3)+A_1$&$\mathfrak{su}(2)$&${\begin{aligned}&(1;3)+(2;4)+4(3;1)+2(4;2)+(5;1)+2(5;3)+3(6;2)\\&+(7;1)+(7;3)+2(8;2)+(9;1)+(9;3)+(10;2)+2(11;1)\end{aligned}}$\\
\hline
$E_7(a_5)$&$\mathfrak{su}(2)$&${\begin{aligned}&(1;3)+6(3;1)+3(4;2)+4(5;1)+3(6;2)+5(7;1)\\&+2(8;2)+3(9;1)+(10;2)+3(11;1)\end{aligned}}$\\
\hline
$D_5+A_1$&$\mathfrak{su}(2)^2$&${\begin{aligned}&(1;1,3)+(1;3,1)+(2;2,3)+2(3;1,1)+(4;1,2)+(5;2,2)\\&+(6;1,2)+(7;1,1)+(8;2,1)+(9;1,3)+(10;1,2)+(10;2,1)\\&+(11;1,1)+(11;2,2)+(12;1,2)+(15;1,1)\end{aligned}}$\\
\hline
$E_8(a_7)$&$-$&$4(11)+6(9)+10(7)+10(5)+10(3)$\\
\hline
$A_6$&$\mathfrak{su}(2)^2$&${\begin{aligned}&(1;1,3)+(1;3,1)+(3;1,1)+(3;2,2)+(5;1,3)+(7;1,5)\\&+(7;2,4)+(9;1,3)+(11;1,1)+(11;2,2)+(13;1,3)\end{aligned}}$\\
\hline
$D_6(a_1)$&$\mathfrak{su}(2)^2$&${\begin{aligned}&(1;1,3)+(1;3,1)+2(3;1,1)+(3;2,2)+(4;1,2)+(4;2,1)\\&+(6;1,2)+(6;2,1)+2(7;1,1)+(9;1,1)+(9;2,2)+(10;1,2)\\&+(10;2,1)+2(11;1,1)+(12;1,2)+(12;2,1)+(15;1,1)\end{aligned}}$\\
\hline
$A_6+A_1$&$\mathfrak{su}(2)$&${\begin{aligned}&(1;3)+(2;2)+2(3;1)+(4;2)+(5;3)+(6;4)+(7;5)\\&+(8;4)+(9;3)+(10;2)+(11;1)+(12;2)+(13;3)\end{aligned}}$\\
\hline
$E_7(a_4)$&$\mathfrak{su}(2)$&${\begin{aligned}&(1;3)+(2;2)+4(3;1)+2(4;2)+2(5;1)+(6;2)+3(7;1)\\&+(8;2)+2(9;1)+2(10;2)+4(11;1)+(12;2)+(13;1)+(15;1)\end{aligned}}$\\
\hline
$E_6(a_1)$&$\mathfrak{su}(3)$&${\begin{aligned}&(1;8)+(3;1)+(5;1)+(5;3)+(5;\overline{3})+(7;1)+(9;1)\\&+(9;3)+(9;\overline{3})+2(11;1)+(13;3)+(13;\overline{3})+(15;1)+(17;1)\end{aligned}}$\\
\hline
$D_5+A_2$&$\mathfrak{u}(1)$&${\begin{aligned}&15_0+13_1+13_{-1}+11_{0}+11_{1}+11_{-1}+11_{2}+11_{-2}+11_{3}+11_{-3}\\&+9_{0}+9_{1}+9_{-1}+9_{2}+9_{-2}+7_{0}+7_{1}+7_{-1}+7_{2}+7_{-2}+5_{0}\\&+5_{1}+5_{-1}+5_{3}+5_{-3}+2(3_{0})+3_{1}+3_{-1}+3_{2}+3_{-2}+3_{4}+3_{-4}+1_{0}\end{aligned}}$\\
\hline
$D_6$&$\mathfrak{sp}(2)$&${\begin{aligned}&(1;10)+(3;1)+(6;4)+(7;1)+(10;4)+(11;1)\\&+(11;5)+(15;1)+(16;4)+(19;1)\end{aligned}}$\\
\hline
$E_6$&$\mathfrak{g}_2$&$(1;14)+(3;1)+(9;7)+(11;1)+(15;1)+(17;7)+(23;1)$\\
\hline
$D_7(a_2)$&$\mathfrak{u}(1)$&${\begin{aligned}&1_0+2_1+2_{-1}+2(3_0)+4_1+4_{-1}+5_0+5_2+5_{-2}\\&+6_1+6_{-1}+3(7_0)+2(8_1)+2(8_{-1})+9_0+9_2+9_{-2}\\&+10_1+10_{-1}+2(11_0)+12_1+12_{-1}+13_0+14_1+14_{-1}+15_0\end{aligned}}$\\
\hline
$A_7$&$\mathfrak{su}(2)$&${\begin{aligned}&(1;3)+(3;1)+(4;2)+(5;3)+(6;2)+(7;1)+(8;4)+(9;3)+(10;2)\\&+(11;1)+(12;2)+(13;3)+(15;1)+(16;2)\end{aligned}}$\\
\hline
$E_6(a_1)+A_1$&$\mathfrak{u}(1)$&${\begin{aligned}&1_0+2_3+2_{-3}+2(3_0)+4_1+4_{-1}+5_0+5_2+5_{-2}\\&+6_1+6_{-1}+7_0+8_1+8_{-1}+9_0+9_2+9_{-2}+10_1+10_{-1}\\&+2(11_0)+12_1+12_{-1}+13_2+13_{-2}+14_1+14_{-1}+15_0+17_0\end{aligned}}$\\
\hline
$E_7(a_3)$&$\mathfrak{su}(2)$&${\begin{aligned}&(1;3)+(2;2)+2(3;1)+(5;1)+(6;2)+2(7;1)+(9;1)+2(10;2)\\&+3(11;1)+(12;2)+2(15;1)+(16;2)+(17;1)+(19;1)\end{aligned}}$\\
\hline
$E_8(b_6)$&$-$&$1(17)+3(15)+2(13)+6(11)+3(9)+5(7)+4(5)+4(3)$\\
\hline
$D_7(a_1)$&$\mathfrak{u}(1)$&${\begin{aligned}&1_0+2(3_0)+3_2+3_{-2}+5_1+5_{-1}+7_0+7_1+7_{-1}+9_0+9_1+9_{-1}+2(11_0)\\&+11_1+11_{-1}+11_2+11_{-2}+13_0+15_0+15_1+15_{-1}+17_1+17_{-1}+19_0\end{aligned}}$\\
\hline
$E_6+A_1$&$\mathfrak{su}(2)$&${\begin{aligned}&(1;3)+(2;4)+2(3;1)+(8;2)+(9;3)+(10;2)\\&+(11;1)+(15;1)+(16;2)+(17;3)+(18;2)+(23;1)\end{aligned}}$\\
\hline
$E_7(a_2)$&$\mathfrak{su}(2)$&${\begin{aligned}&(1;3)+2(3;1)+(4;2)+(7;1)+(8;2)+(9;1)+(10;2)+2(11;1)\\&+2(15;1)+(16;2)+(17;1)+(18;2)+(19;1)+(23;1)\end{aligned}}$\\
\hline
$E_8(a_6)$&$-$&$2(19)+1(17)+3(15)+3(13)+3(11)+3(9)+5(7)+1(5)+3(3)$\\
\hline
$D_7$&$\mathfrak{su}(2)$&${\begin{aligned}&(23;1)+(22;2)+(19;1)+(16;2)+(15;1)+(13;3)+(12;2)+(11;1)\\&+(10;2)+(7;1)+(4;2)+(3;1)+(1;3)\end{aligned}}$\\
\hline
$E_8(b_5)$&$-$&$1(23)+2(19)+3(17)+3(15)+3(11)+3(9)+2(7)+1(5)+4(3)$\\
\hline
$E_7(a_1)$&$\mathfrak{su}(2)$&${\begin{aligned}&(1;3)+(3;1)+(6;2)+(7;1)+2(11;1)+(12;2)+(15;1)\\&+(16;2)+(17;1)+(19;1)+(22;2)+(23;1)+(27;1)\end{aligned}}$\\
\hline
$E_8(a_5)$&$-$&${\begin{aligned}&2(23)+1(21)+1(19)+1(17)+3(15)+2(13)+4(11)+1(9)+1(7)\\&+1(5)+3(3)\end{aligned}}$\\
\hline
$E_8(b_4)$&$-$&${\begin{aligned}&1(27)+2(23)+1(21)+1(19)+2(17)+2(15)\\&+1(13)+3(11)+2(7)+1(5)+2(3)\end{aligned}}$\\
\hline
$E_7$&$\mathfrak{su}(2)$&${\begin{aligned}&(1;3)+(3;1)+(10;2)+(11;1)+(15;1)+(18;2)\\&+(19;1)+(23;1)+(27;1)+(28;2)+(35;1)\end{aligned}}$\\
\hline
$E_8(a_4)$&$-$&${\begin{aligned}&1(29)+1(27)+2(23)+2(19)+1(17)\\&+3(15)+2(11)+1(9)+1(7)+1(5)+1(3)\end{aligned}}$\\
\hline
$E_8(a_3)$&$-$&$1(35)+1(29)+2(27)+1(23)+2(19)+1(17)+1(15)+2(11)+1(9)+2(3)$\\
\hline
$E_8(a_2)$&$-$&${\begin{aligned}&1(39)+1(35)+1(29)+1(27)+2(23)+1(19)+1(17)+1(15)+1(11)\\&+1(7)+1(3)\end{aligned}}$\\
\hline
$E_8(a_1)$&$-$&$1(47)+1(39)+1(35)+1(29)+1(27)+1(23)+1(19)+1(15)+1(11)+1(3)$\\
\hline
$E_8$&$-$&$1(59)+1(47)+1(39)+1(35)+1(27)+1(23)+1(15)+1(3)$\\
\hline
\end{longtable}
}

\section{Projection Matrices}\label{projection_matrices}

{

\begin{longtable}{|c|c|c|}
\hline
Bala-Carter&$\mathfrak{f}$&Projection Matrix\\
\hline
\endhead
$A_1$&$\mathfrak{e}_7$&$\begin{pmatrix}2&4&6&5&4&3&2&3\\1&0&0&0&0&0&0&0 \\ 0&1&0&0&0&0&0&0\\0&0&1&0&0&0&0&0\\0&0&0&1&0&0&0&0\\0&0&0&0&1&0&0&0\\0&0&0&0&0&1&0&0\\0&0&0&0&0&0&0&1\end{pmatrix}$\\
\hline
$2A_1$&$\mathfrak{so}(13)$&$\begin{pmatrix}4&7&10&8&6&4&2&5\\0&0&0&0&0&0&1&0\\0&0&0&0&0&1&0&0\\0&0&0&0&1&0&0&0\\0&0&0&1&0&0&0&0\\0&0&1&0&0&0&0&0\\0&1&0&0&0&0&0&1\end{pmatrix}$\\
\hline
$3A_1$&$\mathfrak{f}_4\times \mathfrak{su}(2)$&$\begin{pmatrix}4&8&12&10&8&6&3&6 \\ 0&0&0&0&0&0&0&1 \\ 0&0&1&0&0&0&0&0 \\ 0&1&0&1&0&0&0&0 \\ 1&0&0&0&1&0&0&0 \\ 0&0&0&0&0&0&1&0\end{pmatrix}$\\
\hline
$A_2$&$\mathfrak{e}_6$&$\begin{pmatrix}4&8&12&10&8&6&4&6 \\ 0&0&0&0&1&1&0&0\\ 0&0&1&1&0&0&0&0 \\ 0&1&0&0&0&0&0&0 \\ 0&0&1&0&0&0&0&1 \\ 0&0&0&1&1&0&0&0 \\ 1&0&0&0&0&0&0&0\end{pmatrix}$\\
\hline
$4A_1$&$\mathfrak{sp}(4)$&$\begin{pmatrix}5&10&15&12&9&6&3&8 \\ 1&0&0&0&0&0&1&0 \\ 0&1&0&0&0&1&0&0 \\ 0&0&1&0&1&0&0&0 \\ 0&0&0&1&0&0&0&0\end{pmatrix}$\\
\hline
$A_2+A_1$&$\mathfrak{su}(6)$&$\begin{pmatrix}6 & 11 & 16 & 13 & 10 & 7 & 4 & 8 \\0&0&0&0&1&1&0&0 \\0&0&1&1&0&0&0&0 \\0&1&0&0&0&0&0&0 \\0&0&1&0&0&0&0&1 \\0&0&0&1&1&0&0&0\end{pmatrix}$\\
\hline
$A_2+2A_1$&$\mathfrak{so}(7)\times \mathfrak{su}(2)$&$\begin{pmatrix}6 & 12 & 18 & 15 & 12 & 8 & 4 & 9 \\1 & 0 & 0&0&0&0&0&0 \\0&1&0&0&0&0&0&0 \\0&0&2&1&0&0&0&1 \\0&0&0&1&0&2&2&1\end{pmatrix}$\\
\hline
$A_3$&$\mathfrak{so}(11)$&$\begin{pmatrix}8&15&22&18&14&10&6&11 \\ 0&0&0&0&1&0&0&0 \\ 0&0&0&1&0&0&0&0 \\ 0&0&1&0&0&0&0&0 \\ 0&0&0&0&0&0&0&1 \\ 0&1&0&0&0&0&0&-1\end{pmatrix}$\\
\hline
$A_2+3A_1$&$\mathfrak{g}_2\times \mathfrak{su}(2)$&$\begin{pmatrix}7&14&20&16&12&8&4&10 \\0&0&0&2&2&0&1&1 \\0&0&1&0&0&1&0&0 \\1&0&0&0&0&0&0&0\end{pmatrix}$\\
\hline
$2A_2$&${\mathfrak{g}}_2^2$&$\begin{pmatrix}8 & 14 & 20 & 16 & 12 & 8 & 4 & 10 \\0&1&2&0&0&1&1&1 \\0&0&0&1&1&0&0&0 \\0&1&0&0&1&1&0&1 \\0&0&1&1&0&0&0&0\end{pmatrix}$\\
\hline
$2A_2+A_1$&$\mathfrak{g}_2 \times \mathfrak{su}(2)$&$\begin{pmatrix}8&15&22&18&14&10&5&11 \\0&1&2&0&1&0&0&1 \\0&0&0&1&0&0&0&0 \\0&1&0&0&0&0&1&1\end{pmatrix}$\\
\hline
$A_3+A_1$&$\mathfrak{so}(7)\times \mathfrak{su}(2)$&$\begin{pmatrix}8&16&24&20&16&11&6&12 \\0&0&0&1&0&0&0&0 \\0&0&1&0&0&0&0&0 \\1&1&0&0&0&0&0&1 \\0&0&0&0&0&1&0&0\end{pmatrix}$\\
\hline
$D_4(a_1)$&$\mathfrak{so}(8)$&$\begin{pmatrix}8 & 16 & 24 & 20 & 16 & 12 & 6 & 12 \\0&0&0&0&1&0&0&0 \\0&0&0&1&0&0&0&0 \\0&0&1&0&0&0&0&0 \\1&1&1&0&0&0&0&1\end{pmatrix}$\\
\hline
$D_4$&$\mathfrak{f}_4$&$\begin{pmatrix}12 & 24 & 36 & 30 & 24 & 18 & 10 & 18 \\0&0&0&0&0&0&0&1 \\0&0&1&0&0&0&0&0 \\0&1&0&1&0&0&0&0 \\1&0&0&0&1&0&0&0\end{pmatrix}$\\
\hline
$2A_2+2A_1$&$\mathfrak{sp}(2)$&$\begin{pmatrix}8 & 16 & 24 & 20 & 15 & 10 & 5 & 12 \\0&2&2&0&1&0&1&0 \\1&0&0&0&0&1&0&1\end{pmatrix}$\\
\hline
$A_3+2A_1$&$\mathfrak{sp}(2)\times \mathfrak{su}(2)$&$\begin{pmatrix}9&18&26&21&16&11&6&13 \\0&0&0&1&0&0&0&1\\0&0&1&0&0&0&0&0 \\1&0&0&0&0&1&0&0\end{pmatrix}$\\
\hline
$D_4(a_1)+A_1$&${\mathfrak{su}(2)}^3$&$\begin{pmatrix}9&18&27&22&17&12&6&14 \\1 &0&0&0&0&0&0&0 \\0&0&1&0&0&0&0&0 \\0&0&0&0&1&0&0&0\end{pmatrix}$\\
\hline
$A_3+A_2$&$\mathfrak{sp}(2)\times\mathfrak{u}(1)$&$\begin{pmatrix}10&19&28&23&18&12&6&14\\0&1&0&1&0&0&0&0 \\0&0&1&0&0&0&0&1\\0&1/2&1&1/2&0&1&1&0\end{pmatrix}$\\
\hline
$A_4$&$\mathfrak{su}(5)$&$\begin{pmatrix}12 & 22 & 32 & 26 & 20 & 14 & 8 & 16\\0&0&0&0&0&0&0&1 \\0&0&1&0&0&0&0&0 \\0&0&0&1&0&0&0&0 \\0&0&0&0&1&0&0&0\end{pmatrix}$\\
\hline
$A_3+A_2+A_1$&${\mathfrak{su}(2)}^2$&$\begin{pmatrix}10&20&30&24&18&12&6&15\\2&2&0&4&6&6&4&0\\0&0&0&0&0&0&0&1\end{pmatrix}$\\
\hline
$D_4+A_1$&$\mathfrak{sp}(3)$&$\begin{pmatrix}13 & 26 & 39 & 32 & 25 & 18 & 10 & 20 \\1 & 0&0&0&1&0&0&0 \\0&1&0&1&0&0&0&0\\0&0&1&0&0&0&0&0\end{pmatrix}$\\
\hline
$D_4(a_1)+A_2$&$\mathfrak{su}(3)$&$\begin{pmatrix}10 & 20 & 30 & 24 & 18 & 12 & 6 & 16 \\1 & 0 & 2 & 2 & 2 & 0 & 1 & 0 \\1 & 3 & 2&2&2&3&1&0\end{pmatrix}$\\
\hline
$A_4+A_1$&$\mathfrak{su}(3)\times \mathfrak{u}(1)$&$\begin{pmatrix}12 & 23 & 34 & 28 & 22 & 15 & 8 & 17 \\0&0&0&0&0&0&0&1 \\0&0&1&0&0&0&0&0 \\0&3&2&0&0&3&0&1\end{pmatrix}$\\
\hline
$2A_3$&$\mathfrak{sp}(2)$&$\begin{pmatrix}12&23&34&28&21&14&7&17\\0&1&0&0&1&0&1&1\\0&0&1&0&0&1&0&0\end{pmatrix}$\\
\hline
$D_5(a_1)$&$\mathfrak{su}(4)$&$\begin{pmatrix}14&27&40&33&26&18&10&20 \\0&0&1&1&0&0&0&0\\0&0&0&0&0&0&0&1\\0&1&1&0&0&0&0&0\end{pmatrix}$\\
\hline
$A_4+2A_1$&$\mathfrak{su}(2)\times \mathfrak{u}(1)$&$\begin{pmatrix}12 & 24 & 36 & 29 & 22 & 15 & 8 & 18 \\0&0&0&1&0&1&0&0 \\2&0&0&1&2&1&0&2\end{pmatrix}$\\
\hline
$A_4+A_2$&${\mathfrak{su}(2)}^2$&$\begin{pmatrix}12 & 24 & 36 & 30 & 24 & 16 & 8 & 18 \\1 & 2 & 2 & 1 & 0 & 0 & 0 & 1 \\1 & 0 & 4 & 3 & 0 & 2 & 2 & 3\end{pmatrix}$\\
\hline
$A_5$&$\mathfrak{g}_2\times\mathfrak{su}(2)$&$\begin{pmatrix}16&30&44&36&28&19&10&22\\0&1&1&0&0&0&0&2\\0&0&0&0&0&0&0&-1\\0&0&0&0&0&1&0&0\end{pmatrix}$\\
\hline
$D_5(a_1)+A_1$&${\mathfrak{su}(2)}^2$&$\begin{pmatrix} 14&28&42&34&26&18&10&21 \\ 0&0&0&0&0&0&0&1 \\ 2&0&0&2&0&0&0&0 \end{pmatrix}$\\
\hline
$A_4+A_2+A_1$&$\mathfrak{su}(2)$&$\begin{pmatrix}13&26&38&31&24&16&8&19\\1&0&0&1&0&2&0&3\end{pmatrix}$\\
\hline
$D_4+A_2$&$\mathfrak{su}(3)$&$\begin{pmatrix} 14&28&42&34&26&18&10&22 \\ 0&1&0&1&0&0&0&0 \\ 1&0&0&0&1&0&0&0 \end{pmatrix}$\\
\hline
$E_6(a_3)$&$\mathfrak{g}_2$&$\begin{pmatrix} 16&30&44&36&28&20&10&22 \\ 0&1&1&1&1&0&0&2 \\ 0&0&0&0&0&0&0&-1 \end{pmatrix}$\\
\hline
$D_5$&$\mathfrak{so}(7)$&$\begin{pmatrix}20&38&56&46&36&26&14&28\\0&0&0&0&0&0&0&1\\0&0&1&0&0&0&0&0\\0&1&0&1&0&0&0&0\end{pmatrix}$\\
\hline
$A_4+A_3$&$\mathfrak{su}(2)$&$\begin{pmatrix}14&28&42&34&26&18&9&21\\2&0&0&2&0&0&1&1\end{pmatrix}$\\
\hline
$A_5+A_1$&${\mathfrak{su}(2)}^2$&$\begin{pmatrix}16&31&46&37&28&19&10&23 \\0&1&0&0&0&1&0&1 \\ 0&0&0&1&0&0&0&0 \end{pmatrix}$\\
\hline
$D_5(a_1)+A_2$&$\mathfrak{su}(2)$&$\begin{pmatrix}15&30&44&36&28&19&10&22\\1&0&2&0&0&1&0&0\end{pmatrix}$\\
\hline
$D_6(a_2)$&${\mathfrak{su}(2)}^2$&$\begin{pmatrix}16&32&47&38&29&20&10&24 \\ 0&0&0&0&1&0&0&0\\ 0&0&1&0&0&0&0&0 \end{pmatrix}$\\
\hline
$E_6(a_3)+A_1$&$\mathfrak{su}(2)$&$\begin{pmatrix} 16&31&46&38&29&20&10&23 \\ 0&1&0&0&1&0&0&1 \end{pmatrix}$\\
\hline
$E_7(a_5)$&$\mathfrak{su}(2)$&$\begin{pmatrix}16&32&48&39&30&20&10&24 \\ 0&0&0&1&0&0&0&0 \end{pmatrix}$\\
\hline
$D_5+A_1$&${\mathfrak{su}(2)}^2$&$\begin{pmatrix}20&39&58&48&37&26&14&29\\0&0&1&0&0&0&0&0\\0&1&1&0&1&0&0&1\end{pmatrix}$\\
\hline
$E_8(a_7)$&$-$&$\begin{pmatrix}16&32&48&40&30&20&10&24 \end{pmatrix}$\\
\hline
$A_6$&${\mathfrak{su}(2)}^2$&$\begin{pmatrix}20&38&56&46&36&24&12&28 \\ 0&0&0&0&0&0&0&1 \\ 0&0&2&2&0&2&2&1 \end{pmatrix}$\\
\hline
$D_6(a_1)$&${\mathfrak{su}(2)}^2$&$\begin{pmatrix}20&40&59&48&37&26&14&30\\0&0&0&0&1&0&0&0\\0&0&1&0&0&0&0&0\end{pmatrix}$\\
\hline
$A_6+A_1$&$\mathfrak{su}(2)$&$\begin{pmatrix}20&39&58&47&36&24&12&29\\0&1&0&1&0&2&2&1\end{pmatrix}$\\
\hline
$E_7(a_4)$&$\mathfrak{su}(2)$&$\begin{pmatrix}20&40&60&49&38&26&14&30\\0&0&0&1&0&0&0&0\end{pmatrix}$\\
\hline
$E_6(a_1)$&$\mathfrak{su}(3)$&$\begin{pmatrix}24&46&68&56&44&30&16&34\\0&0&0&1&0&0&0&0\\0&0&1&0&0&0&0&0\end{pmatrix}$\\
\hline
$D_5+A_2$&$\mathfrak{u}(1)$&$\begin{pmatrix}20&40&60&50&38&26&14&30\\2&0&0&0&1&0&0&1\end{pmatrix}$\\
\hline
$D_6$&$\mathfrak{sp}(2)$&$\begin{pmatrix}28&54&79&64&49&34&18&40\\0&0&1&0&1&0&0&0\\0&0&0&1&0&0&0&0\end{pmatrix}$\\
\hline
$E_6$&$\mathfrak{g}_2$&$\begin{pmatrix}32&62&92&76&60&42&22&46\\0&1&0&1&0&0&0&1\\0&0&1&0&0&0&0&0\end{pmatrix}$\\
\hline
$D_7(a_2)$&$\mathfrak{u}(1)$&$\begin{pmatrix}22&43&64&52&40&27&14&32\\0&1&0&0&0&1&0&0\end{pmatrix}$\\
\hline
$A_7$&$\mathfrak{su}(2)$&$\begin{pmatrix}24&47&70&57&44&30&15&35\\0&1&0&1&0&0&1&1\end{pmatrix}$\\
\hline
$E_6(a_1)+A_1$&$\mathfrak{u}(1)$&$\begin{pmatrix}24&47&70&57&44&30&16&35\\0&1&0&1&0&0&0&1\end{pmatrix}$\\
\hline
$E_7(a_3)$&$\mathfrak{su}(2)$&$\begin{pmatrix}28&54&80&65&50&34&18&40\\0&0&0&1&0&0&0&0\end{pmatrix}$\\
\hline
$E_8(b_6)$&$-$&$\begin{pmatrix}24&48&72&58&44&30&16&36\end{pmatrix}$\\
\hline
$D_7(a_1)$&$\mathfrak{u}(1)$&$\begin{pmatrix}28&54&80&66&50&34&18&40\\0&0&0&0&0&1&0&1\end{pmatrix}$\\
\hline
$E_6+A_1$&$\mathfrak{su}(2)$&$\begin{pmatrix}32&63&94&77&60&42&22&47\\0&1&0&1&0&0&0&1\end{pmatrix}$\\
\hline
$E_7(a_2)$&$\mathfrak{su}(2)$&$\begin{pmatrix}32&64&95&78&60&42&22&48\\0&0&1&0&0&0&0&0\end{pmatrix}$\\
\hline
$E_8(a_6)$&$-$&$\begin{pmatrix}28&56&84&68&52&36&18&42\end{pmatrix}$\\
\hline
$D_7$&$\mathfrak{su}(2)$&$\begin{pmatrix}36&70&103&84&64&43&22&52\\0&0&1&0&0&1&0&0\end{pmatrix}$\\
\hline
$E_8(b_5)$&$-$&$\begin{pmatrix}32&64&96&78&60&42&22&48\end{pmatrix}$\\
\hline
$E_7(a_1)$&$\mathfrak{su}(2)$&$\begin{pmatrix}40&78&115&94&72&50&26&58 \\ 0&0&1&0&0&0&0&0\end{pmatrix}$\\
\hline
$E_8(a_5)$&$-$&$\begin{pmatrix}36&70&104&84&64&44&22&52\end{pmatrix}$\\
\hline
$E_8(b_4)$&$-$&$\begin{pmatrix}40&78&116&94&72&50&26&58\end{pmatrix}$\\
\hline
$E_7$&$\mathfrak{su}(2)$&$\begin{pmatrix}52&102&151&124&96&66&34&76\\0&0&1&0&0&0&0&0\end{pmatrix}$\\
\hline
$E_8(a_4)$&$-$&$\begin{pmatrix}44&86&128&104&80&54&28&64\end{pmatrix}$\\
\hline
$E_8(a_3)$&$-$&$\begin{pmatrix}52&102&152&124&96&66&34&76\end{pmatrix}$\\
\hline
$E_8(a_2)$&$-$&$\begin{pmatrix}60&118&174&142&108&74&38&88\end{pmatrix}$\\
\hline
$E_8(a_1)$&$-$&$\begin{pmatrix}72&142&210&172&132&90&46&106\end{pmatrix}$\\
\hline
$E_8$&$-$&$\begin{pmatrix}92&182&270&220&168&114&58&136\end{pmatrix}$\\
\hline
\end{longtable}
}

\end{appendices}

\bibliographystyle{utphys}

%\small\baselineskip=.93\baselineskip
%\let\bbb\bibitem\def\bibitem{\itemsep1pt\bbb}
\bibliography{ref}

\end{document}